\documentclass[pra, twocolumn,showpacs,nofootinbibfloatfix,amsmath,amsfonts,amssymb]{revtex4-1}%
\usepackage{amsmath,amsfonts,amssymb,color}
\usepackage{amsthm}
\usepackage{leftidx}
\usepackage{graphicx}
\usepackage{xcolor}
\usepackage{dcolumn}
\usepackage{bm}
\usepackage{epstopdf}
\usepackage{epsfig}
\usepackage{mathdots} 

\begin{document}
	\title{Time-induced second-order topological superconductors}
	\author{Raditya Weda Bomantara}
	\email{Raditya.Bomantara@sydney.edu.au}
	\affiliation{%
		Centre for Engineered Quantum Systems, School of Physics, University of Sydney, Sydney, New South Wales 2006, Australia
	}
	\date{\today}
	
	
	\vspace{2cm}
	
\begin{abstract}
Higher-order topological materials with topologically protected states at the boundaries of their boundaries (hinges or corners) have attracted attention in recent years. In this paper, we utilize time-periodic driving to generate second-order topological superconductors out of systems which otherwise do not even allow second-order topological characterization. This is made possible by the design of the periodic drives which inherently exhibit nontrival winding in the time-domain. Through the interplay of topology in both spatial and temporal dimensions, nonchiral Majorana modes may emerge at the systems' corners and sometimes even coexist with chiral Majorana modes. Our proposal thus presents a unique opportunity to Floquet engineering with minimal system's complexity and its application in quantum information processing. 
	
\end{abstract}

\maketitle

\section{Introduction} 
\label{intro}

Since their theoretical discoveries in early 1980s \cite{Thou1,Thou2}, followed by various experimental realizations since the last decade \cite{expt1,expt2}, topological phases of matter have remained an active field of research. Their main signature to host robust topologically protected states in the presence of systems' boundaries or defects is especially attractive with potential applications in designing robust electronic/spintronic devices \cite{app1} and fault-tolerant quantum computing \cite{app2,app3}. 

In the last couple of years, a new direction within the area of topological matter emerges through the discovery of higher-order topological phases (HOTP) \cite{HTI-1,HTI0,HTI1,HTI2,HTI3}, which exhibit topologically protected states at the boundaries of the systems' boundaries. In particular, an $n$-th-order topological phase in $D$ dimensions is characterised by the presence of topologically protected states at its $D-n$ dimensional boundaries. In the following years, HOTP have been extensively studied \cite{HTI10,HTI11,HTI12,HTI12b,HTI12c,HTI13,HTI14,HTI15,HTI16,HTI17,HTI18,HTI19,HTI20,HTI21,HTI22,HTI23,HTI24,HTI4,HTI5,HTI6,HTI7,HTI8,HTI9,HTI25,HTI26} and experimentally observed in a variety of physical platforms, such as photonics \cite{HTI19,HTI24},  acoustics \cite{HTI25}, electrical circuit devices \cite{HTI20}, and solid-state systems \cite{HTI18}. A common feature of these studies suggests that systems with at least four bands are necessary for the formation of HOTP. As a result, a construction of such HOTP inevitably requires a number of internal degrees of freedom and/or spatial variations in the system parameters, thus leading to a generally complex design.   


In a slightly different aspect, the possibility of using periodic drives to generate nontrivial topology in an otherwise trivial static system has resulted in various studies of Floquet topological phases since the last decade \cite{Flor1,Rud,Flor10,Flor11,Flor12,Flor13,Flor14,Flor15,Flor16,Flor17,Flor18,Flor19,Flor2,Flor20,Flor21,Flor22,Long,Flor23,Flor24,Flor3,Flor32,Flor4,Flor5,Flor6,Flor7,Flor8,FMF1,FMF2,FMF3,FMF4,FMF5,FMF6,RG,RG2,FHTI1,FHTI2,FHTI3,FHTI4,FHTI5,FHTI6,FHTI7,YP1,YP2,YP3}. In such time-periodic systems, energy is no longer a conserved quantity and is replaced by the so-called quasienergy, which is only defined modulo the frequency of the drive. Such a periodicity of quasienergies leads to the existence of an additional gap (termed Floquet zone-edge gap \cite{Rud}) which allows the emergence of topological features with no static counterparts, such as chiral \cite{Flor2,Flor3,Flor32,Flor4,Flor5,Flor6,Flor7,Flor8,Rud} or dispersionless \cite{FMF1,FMF2,FMF3,FMF4,FMF5,FMF6,RG,RG2,Rud} edge states at the Floquet zone-edge gap. As rigorously studied in Ref.~\cite{Rud}, it is so far understood that characterization of these features involves the interplay between known topological invariants in static systems (e.g., the Chern numbers) and an additional invariant unique to Floquet systems characterising the topology of the Floquet zone-edge gap. 

This paper aims to take a step forward in the aforementioned directions (i.e., Floquet topological phases and HOTP) by directly equipping periodic drives themselves with nontrivial topology, which allows the emergence of HOTP in the resulting driven system even when the underlying static system does not support such a topological characterization. { Here, topology of the drives (which we refer to as time-induced topology) simply refers to the winding number made by the quantity $h_c(t)+\mathrm{i} h_s(t)$ in the time-domain, where $h_c(t)$ and $h_s(t)$ are two time-periodic terms in the system's Hamiltonian. In practice, it can actually be very easily implemented by properly introducing two harmonic drives with the same frequency $\omega$ and a relative phase difference of $\pi/2$ (i.e., $h_c(t)\propto \cos(\omega t)$ and $h_s(t)\propto \sin(\omega t)$),} which offers an important advantage of significantly reducing the required systems' complexity for hosting HOTP or potentially other exotic topological phases.

While the idea presented in this paper can be generalized to other HOTP, we focus on the generation of second-order topological superconductors (SOTSCs) from an inherently trivial two-band $p_x+\mathrm{i}p_y$ superconductor due to the former's ability to host non-chiral Majorana modes (MMs). Such non-chiral MMs are particularly attractive due to their role as building blocks of nonlocal qubits in topological quantum computing \cite{app2,app3}. These non-chiral MMs are usually found at the ends of certain one-dimensional (1D) systems, i.e., first-order topological superconductors. As a result, the implementation of quantum gate operations, accomplished by moving some MMs around one another (a process termed braiding), generally requires the design of complex branched architectures \cite{braid1,braid2,braid4} which may additionally pose technical challenges \cite{bprob}. 

In recent years, more sophisticated architectures based on arrays of nanowires and measurement-based braiding are proposed \cite{mr1,braid5} to avoid the use of any branched structures. A minimal model of such architectures, termed a tetron in \cite{mr1} or a Majorana cooper-pair box (MCB) in Ref.~\cite{braid5}, has been considered as a promising building block for Majorana-based surface codes \cite{mr2,mr3,mr4,surcode} and larger-scale qubit architectures. In practice, however, MCBs or tetrons based on two parallel nanowires may suffer from unequal charging energies and mutual capacitive coupling between them \cite{mr1,mr3}, which may be detrimental for unleashing their full potential. On the other hand, a single two-dimensional (2D) first-order topological superconductor may only host chiral MMs at its edges \cite{CMM,CMM2}, which are not directly relevant for quantum computing applications. While non-chiral MMs may also appear at the vortices in the bulk of certain 2D fractional quantum Hall systems \cite{mmvort}, the latter is challenging to realize experimentally, and such MMs are generally fixed in place and may not be readily manipulated to perform quantum gate operations. 

For the above reasons, realizing non-chiral MMs with 2D SOTSCs is especially advantageous not only because these MMs naturally exist without the introduction of vortices, but also that braiding of MMs can be more realistically implemented either through conductance-measurements \cite{RG} or adiabatic following \cite{adiac1,adiac2,adiac3,adiac4}. A single SOTSC also naturally forms a tetron/MCB of Refs.~\cite{mr1,braid5} with uniform charging energy across all MMs and without the introduction of mutual capacitive coupling elucidated above. Moreover, due to the possible coexistence of two species of non chiral MMs termed Majorana zero modes (MZMs) and Majorana $\pi$ modes (MPMs), both of which are capable of encoding qubits \cite{RG}, Floquet SOTSC-based tetrons/MCBs potentially offer the additional advantage of significantly reducing the physical resources for designing a given topological qubit architecture. A potential drawback of such Floquet SOTSC-based tetrons/MCBs currently lies in the design of the SOTSC itself, which as elucidated before necessarily requires spatial variations of some system parameters and/or additional degrees of freedom for enabling second-order topology. In this case, the time-induced SOTSC proposal introduced in this work overcomes this weakness, thus increasing the attractiveness of SOTSC-based qubit architectures for near future experiments.       

This paper is structured as follows. In Sec.~\ref{model}, we present a minimal model demonstrating the possibility of encoding topology in the time-domain and briefly review Floquet theory. In Sec.~\ref{symprot}, we introduce a set of infinite matrices with algebra similar to that of $2\times 2$ Pauli matrices, which allow the characterization of the system's symmetries. We then elucidate how these symmetries allow the characterisation of the system's whole topology by inspecting only a diagonal and anti-diagonal line in the 2D Brillouin zone. In Sec.~\ref{appr}, we explicitly derive a $Z_2$ invariant predicting the existence of corner MPMs in the system. In Sec.~\ref{exat}, we present our numerical calculations which explicitly verify the presence of these corner MPMs. In Sec.~\ref{coexist}, we highlight a rare scenario in which chiral and non-chiral MMs coexist, as well as its potential application for transferring Majorana-based quantum information. We further compare our work with previous literature. Finally, we summarize the paper and highlight opportunities for potential future studies in Sec.~\ref{conc}.    

\section{Time-induced topology}

\subsection{Minimal model}
\label{model}

To illustrate the main physics, we consider a (2D) square lattice model describing a periodically driven $p_x+i p_y$ superconductor,
\begin{eqnarray}
H(t) &=& \sum_{i,j} \left[\mu c_{i,j}^\dagger c_{i,j}+ \left( J_x(t) c_{i+1,j}^\dagger c_{i,j} +  J_y(t) c_{i,j+1}^\dagger c_{i,j}\right. \right. \nonumber \\ 
&& \left. \left. + \Delta c_{i+1,j}^\dagger c_{i,j}^\dagger + \mathrm{i} \Delta  c_{i,j+1}^\dagger c_{i,j}^\dagger +h.c.\right) \right] \;,
\label{sys}
\end{eqnarray}
where $c_{i,j}^\dagger$ ($c_{i,j}$) is the fermionic creation (annihilation) operator at lattice site $(i,j)$, $\mu$ represents the chemical potential, $J_x(t)=J_{s,x}+J_{0,x}\cos(\omega t)$ and $J_y(t) = J_{s,y}+J_{0,y}\sin(\omega t)$ are the time-periodic hopping amplitudes of period $T=\frac{2\pi}{\omega}$ in the $x$ and $y$ directions respectively, and $\Delta \in \mathbb{R}$ characterizes the $p_x+\mathrm{i} p_y$ pairing strength. While Eq.~(\ref{sys}) looks like a toy model, its static version has actually been experimentally realized in Ref.~\cite{CMM} to detect the existence of chiral MMs, where effective $p_x+\mathrm{i} p_y$ superconductivity is realized by proximitizing a quantum anomalous Hall insulator thin film, such as (Cr$_{0.12}$Bi$_{0.26}$Sb$_{0.62}$)$_2$Te$_3$, with a normal ($s$-wave) superconductor. Within this framework, the effective $p_x+\mathrm{i} p_y$ pairing is proportional to the fermi velocity of the thin film's top and bottom surface surfaces, the chemical potential is renormalized by the $s$-wave pairing, and the hopping amplitudes are related to the hybridization between the thin film's top and bottom surface states \cite{CMM,ctscprop1,ctscprop2}. Such a hybridization depends on the distance between the two surface states, which can therefore be controlled by either varying the thickness of the thin film or the localization length of the surface states (which can indirectly be achieved by controlling the thin film's band structure). For the purpose of realizing the time-dependence of the hopping amplitudes above, the latter approach is expected to be more feasible. For example, by realizing that such a thin film is the 2D limit of a 3D topological insulator (TI) \cite{d2d3}, well-known driving mechanisms for generating 3D TIs with tunable band gap (such as via electromagnetic radiation \cite{Flor32}) can in principle be employed.

Since Eq.~(\ref{sys}) is time-periodic, we may employ Floquet theory \cite{Flo1,Flo2}. To this end, we construct a Floquet Hamiltonian in an enlarged (Sambe) Hilbert space defined as
\begin{eqnarray}
\left[\mathcal{H}_{\alpha \beta}\right]_{ab} &=& a \hbar \omega \delta_{a,b} \delta_{\alpha,\beta} + H_{\alpha \beta,ab} \;,
\label{Sambe}  
\end{eqnarray}  
where $\alpha$ and $\beta$ are integers running through the dimension of $H(t)$, $a$ and $b$ are the photon indices, i.e., integers running from $-\infty$ to $+\infty$, and $H_{\alpha \beta, ab}=\frac{1}{T}\int_0^T dt \; H_{\alpha \beta}(t) e^{-\mathrm{i} (a-b)\omega t}$. It is noted that $\mathcal{H}_{\alpha \beta}$ is of infinite dimension and, consequently, has an infinite number of eigenvalues (termed quasienergies). However, two quasienergies $\varepsilon$ and $\varepsilon+ \hbar \omega$ describe the same physical states \cite{Flo1,Flo2}. As such, it is sufficient to restrict our attention within the first quasienergy Brillouin Zone $\left(-\frac{\hbar \omega}{2}, \frac{\hbar \omega}{2}\right]$. 

Similar to its static counterpart, the Floquet Hamiltonian $\mathcal{H}_{\alpha \beta}$ may admit Hermitian excitations with $\varepsilon=0$, usually referred to as Majorana zero modes (MZMs). These MZMs commute with $\mathcal{H}_{\alpha \beta}$ and lead to all its quasienergies being at least two-fold degenerate. Due to the periodicity of quasienergy Brillouin Zone, however, Hermitian excitations with $\varepsilon=\frac{\hbar \omega}{2}$ (termed Majorana $\pi$ modes (MPMs) \cite{FMF1,FMF2,FMF3,FMF4,FMF5,RG}) are also allowed. Such MPMs are unique to Floquet systems and lead to all quasienergies of $\mathcal{H}_{\alpha \beta}$ exhibiting $\hbar \omega/2$ spacing. 

Under periodic boundary conditions (PBC), Eq.~(\ref{sys}) can be recast in terms of quasimomenta $k_x$ and $k_y$ as
\begin{eqnarray}
H(t) &=& \sum_{k_x,k_y} \frac{1}{2} \Psi^\dagger_k h_{BdG} \Psi_k \;, \nonumber \\
h_{\rm BdG}(t) &=& h_{0,\rm BdG} + 2h_{c,\rm BdG}\cos(\omega t)+2h_{s,\rm BdG} \sin(\omega t) \;, \nonumber \\
\end{eqnarray} 
where $h_{\rm BdG}$ is the momentum space Bogoliubov-de-Gennes  Hamiltonian, $\Psi_k=\left(c_k, c_{-k}^\dagger\right)^T$ is the Nambu wave function, $\sigma_i$'s are Pauli matrices acting in this Nambu basis, and
\begin{eqnarray}
 h_{0,\rm BdG} &  =&  2 \Delta \sin(k_x) \sigma_y +2 \Delta \sin(k_y) \sigma_x \nonumber\\
	&&  +\left[\mu+J_{s,x}\cos(k_x)+J_{s,y}\cos(k_y)\right]  \sigma_z \;, \nonumber \\
h_{c,\rm BdG}&=& J_{0,x} \cos(k_x) \sigma_z \;, \nonumber \\
h_{s,\rm BdG} &=& J_{0,y} \cos(k_y) \sigma_z \;.
\end{eqnarray} 
The momentum space Floquet Hamiltonian associated with $h_{\rm BdG}$ is then obtained as
\begin{equation}
\mathcal{H}_{\rm BdG} = h_{\rm 0,BdG}\xi_0 + \frac{\hbar \omega}{2} (\sigma_0 \xi_0+\xi_z) + h_{\rm c, BdG} \xi_x + h_{\rm s, BdG} \xi_y
\label{FLOH}
\end{equation}
where $\sigma_0$ is the identity $2\times 2$ matrix and $\xi_i$'s are infinite dimensional matrices representing the Floquet photon indices with elements
\begin{eqnarray}
[\xi_0]_{ab} &=& \delta_{a,b}\;, \nonumber \\
\;[\xi_x]_{ab} &=& \delta_{a,b+1}+\delta_{a,b-1}\;, \nonumber \\
\;[\xi_y]_{ab} &=& \mathrm{i}\;\left(\delta_{a,b+1}-\delta_{a,b-1}\right) \;, \nonumber \\
\;[\xi_z]_{ab} &=& (2b-1) \delta_{a,b} \;, \label{infp}
\end{eqnarray}
$a$ and $b$ are photon indices running from $-\infty$ to $\infty$.

\subsection{Symmetries protection}
\label{symprot}

It is first noted that while $\xi_i$'s are defined such that they look like the generalization of Pauli matrices in infinity dimensions, they do not satisfy the same algebra as the $2\times 2$ Pauli matrices, e.g., $\xi_x$ and $\xi_y$ commute instead of anticommute. However, we can define another set of infinite matrices $\eta_i$'s with elements  
\begin{eqnarray}
[\eta_x(\phi)]_{ab} &=& \exp\left[\mathrm{i} (2b-1)\phi\right]\delta_{1-a,b}\;, \nonumber \\
\;[\eta_y(\phi)]_{ab} &=& \exp\left[\mathrm{i} (2b-1)(\phi-\pi/2)\right]\delta_{1-a,b} \;, \nonumber \\
\;[\eta_z]_{ab} &=& (-1)^b \delta_{a,b} \;, \label{infp2}
\end{eqnarray}
where $\phi\in[0,2\pi)$. It can be verified that $\eta_x$, $\eta_y$, and $\eta_z$ are mutually anticommuting, and they transform as $\eta_i\eta_j=\delta_{i,j}+\mathrm{i}\epsilon_{ijk}\eta_k$ similar to $2\times 2$ Pauli matrices. Moreover, for $\xi_\phi=\cos(\phi)\xi_x+\sin(\phi)\xi_y$, we have $\eta_x \xi_\phi \eta_x= -\eta_y \xi_\phi \eta_y=-\eta_z \xi_\phi \eta_z=\xi_\phi$ and $\eta_i \xi_z \eta_i=(2\delta_{i,z}-1)\xi_z$, so that $\eta_i$'s interact with $\xi_i$'s as if they are the same set of Pauli matrices. 

Using the generalized Pauli matrices $\eta_i$'s at a specifically chosen $\phi=\arctan\left(\frac{J_{0,y}}{J_{0,x}}\right)$, we may now identify the system's symmetries similar to the way it is usually done in static systems. Namely, there exists a particle-hole symmetry as well as diagonal and anti-diagonal spatial symmetries \cite{note0} about quasienergy $\frac{\hbar\omega}{2}$, which satisfy (respectively)
\begin{eqnarray}
\mathcal{P} \tilde{\mathcal{H}}(\mathbf{k}) \mathcal{P}^{-1} &=& -\tilde{\mathcal{H}}(-\mathbf{k}) \;, \nonumber \\
\mathcal{M}_{\rm D}\tilde{\mathcal{H}}(k_x=k_y)\mathcal{M}_{\rm D}^{-1}&=&-\tilde{\mathcal{H}}(k_x=k_y) \;, \nonumber \\
\mathcal{M}_{\rm AD}\tilde{\mathcal{H}}(k_x=-k_y)\mathcal{M}_{\rm AD}^{-1}&=&-\tilde{\mathcal{H}}(k_x=-k_y)\;, \nonumber \\ \label{sym0}
\end{eqnarray}
where $\mathcal{P}=\sigma_x \eta_x(0) \mathcal{K}$,
\begin{eqnarray}
\tilde{\mathcal{H}}(\mathbf{k})&=&\mathcal{H}_{\rm BdG}(\mathbf{k})-\frac{\hbar \omega}{2} \sigma_0\xi_0 \;, \nonumber \\ 
\mathcal{M}_{\rm D}&=&\frac{1}{\sqrt{2}}(\sigma_x-\sigma_y)\eta_x(\phi) \;, \nonumber \\
\mathcal{M}_{\rm AD} &=& \frac{1}{\sqrt{2}}(\sigma_x+\sigma_y)\eta_x(\phi) \;, \label{sym} 
\end{eqnarray}
and $\mathcal{K}$ is the complex conjugate. { By defining another infinite matrix $\left[\tilde{\eta}_x\right]_{a,b}=\delta_{a,-b}$, one may also identify the second particle hole symmetry $\tilde{\mathcal{P}}=\sigma_x\tilde{\eta}_x\mathcal{K}$ about quasienergy zero. It maps $\tilde{\mathcal{P}}\mathcal{H}_{\rm BdG}(\mathbf{k})\tilde{\mathcal{P}}^{-1}=-\mathcal{H}_{\rm BdG}(-\mathbf{k})$.} In this case, both particle-hole symmetries are responsible to protect MZMs and MPMs \cite{note}, whereas the two spatial symmetries guarantee that such MZMs and MPMs, if exist, must be localized at the system's corners.  

\begin{figure}
	\begin{center}
		\includegraphics[scale=0.5]{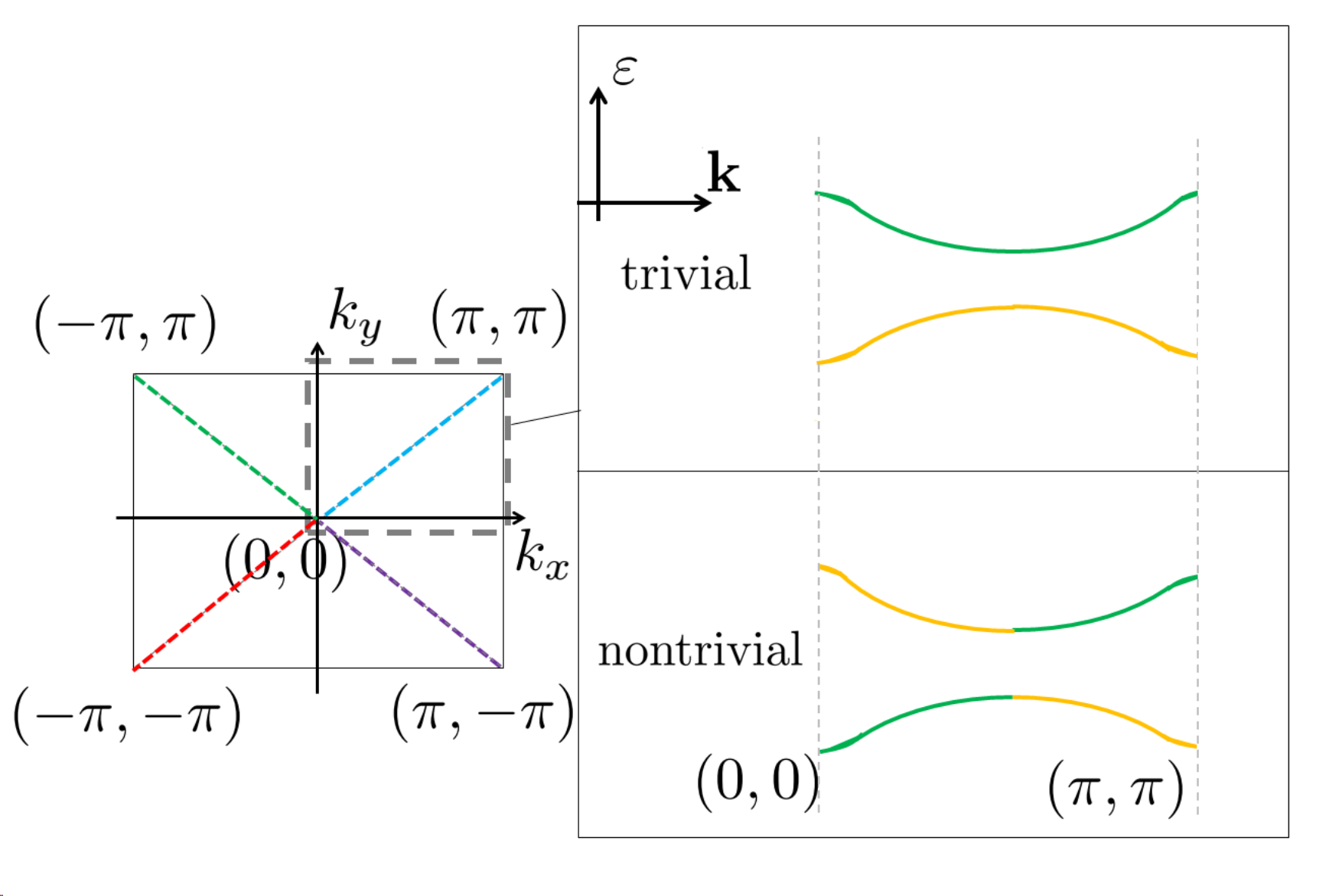}
		\caption{Due to $\mathcal{P}$ and $\tilde{\mathcal{P}}$ symmetries, diagonal (anti-diagonal) line in the 2D Brillouin zone can be further divided into two equivalent subregions marked by the blue and red (green and purple) coloured lines. The inset illustrates two representative many-body quasienergy bands along the blue diagonal line. There, the bands' colours (green and yellow) label the two different eigenstate parities defined in Eq.~(\ref{eigpar}).}
		\label{pic6}
	\end{center}
\end{figure} 

In addition to the four symmetries above, the time-periodic Hamiltonian of Eq.~(\ref{sys}) commutes with the total parity operator, i.e., $P=\prod_{i,j} \mathrm{i} \gamma_{2i,j} \gamma_{2i+1,j}$, where $\gamma_{2i,j}$ and $\gamma_{2i+1,j}$ are two Majorana operators at lattice site $(i,j)$ and are related to the fermionic operator $c_{i,j}$ as
\begin{equation}
c_{i,j}=\frac{1}{2}\left(\gamma_{2i,j}-\mathrm{i}\gamma_{2i+1,j}\right) \;. \label{Majsup2}
\end{equation}
As a result, the system's many-body Floquet eigenstate $|\psi_n\rangle$ also satisfies
\begin{equation}
P|\psi_n\rangle=p_n |\psi_n\rangle \;, \label{eigpar} 
\end{equation}
where $p_n=\pm 1$ is an eigenvalue of $P$ which will be referred to as \emph{eigenstate parity}. { Physically, it can also be understood as the fermion number parity associated with the $n$th many-body quasienergy band.}


We now further elaborate the interplay of the symmetries above in the characterisation of the system's topology. { To this end, we first note that at parameter values $J_{s,x}=J_{s,y}$ and $J_{0,x}=J_{0,y}$, $\mathcal{M}_{\rm D}$ and $\mathcal{M}_{\rm AD}$ become diagonal and anti-diagonal symmetries respectively, which map
\begin{eqnarray}
\mathcal{M}_{\rm D} \mathcal{H}'_{\rm BdG}(k_x,k_y)\mathcal{M}_{\rm D}^{(-1)}&=&-\mathcal{H}'_{\rm BdG}(k_y,k_x) \;, \nonumber \\
\mathcal{M}_{\rm AD} \mathcal{H}'_{\rm BdG}(k_x,k_y)\mathcal{M}_{\rm AD}^{(-1)}&=&-\mathcal{H}'_{\rm BdG}(-k_y,-k_x) \;. \label{mirror} 
\end{eqnarray}
Consequently, a quasienergy gap closing at some generic point $(k_{x,0},k_{y,0})$ in the 2D Brillouin zone must be accompanied by three additional gap closing points at $(k_{y,0},k_{x,0})$, $(-k_{y,0},-k_{x,0})$, and $(-k_{x,0},-k_{y,0})$ due to $\mathcal{M}_{\rm D}$, $\mathcal{M}_{\rm AD}$, and $\mathcal{M}_{\rm D}\mathcal{M}_{\rm AD}$ respectively. Away from the above parameter values, it is expected that a generalization of Eq.~(\ref{mirror}) exists which relates $\mathcal{H}_{\rm BdG}'$ at $(k_{x,0},k_{y,0})$ with that at three other points $(k_{x,1},k_{y,1})$, $(k_{x,2},k_{y,2})$, and $(k_{x,3},k_{y,3})$, whose exact locations depend on $k_{x,0}$, $k_{y,0}$, $J_{0,x}$, $J_{0,y}$, $J_{s,x}$, and $J_{s,y}$.    

In principle, a gap closing point at $(k_{x,0},k_{y,0})$ can be moved towards a diagonal or anti-diagonal line (if it is not already there) before it subsequently annihilates with one of its $\mathcal{M}_{\rm D}$, $\mathcal{M}_{\rm AD}$, and $\mathcal{M}_{\rm D}\mathcal{M}_{\rm AD}$ symmetric-conjugate partners. Such a gap closing and reopening event is therefore topologically equivalent to that occurring along a diagonal or anti-diagonal line. Moreover, due to $\mathcal{P}$ and $\tilde{\mathcal{P}}$, gap closing points along a diagonal or anti-diagonal line must further come in pairs, which are related by a reflection around $(0,0)$.


Combining the two mechanisms above, it follows that the system's topology can be characterised solely from the many-body Floquet bands' properties along a half diagonal and anti-diagonal line as illustrated in Fig.~\ref{pic6}. In particular, consider the restriction of the many-body Floquet bands along the half diagonal line ending at $(0,0)$ and $(\pm \pi, \pm \pi)$. In this case, a single gap closing and reopening event introduces a twist in these bands' eigenstate parity structure along the line. In general, such a twist may occur at any point along the line. However, in the system under our study, it typically emerges at $(\pi/2,\pi/2)$, as Eq.~(\ref{FLOH}) suggests that this is the location at which the gap around $\frac{\hbar \omega}{2}$ quasienergy excitation closes and reopens. 

Depending on the number of such twists, the bands at $(0,0)$ and $(\pm \pi, \pm \pi)$ may have the same or opposite eigenstate parities. The relative eigenstate parity between these two end points thus serves as an invariant characterising two topologically distinct regimes. In the inset of Fig. 1, we illustrate the system's two representative many-body Floquet bands along the blue dashed line in the topologically trivial and nontrivial regime. There, the two possible eigenstate parity values $\pm 1$ are marked by the yellow and green colours of the associated bands. In the topologically trivial (nontrivial) regime, the bands at $(0,0)$ and $(\pi,\pi)$ have the same (opposite) eigenstate parities and are thus marked by the same (different) colours. Physically, a topological nontrivial regime is marked by the presence of MPMs and/or MZMs when the system admits open boundary conditions (OBC). In particular, when nontrivial twists in the eigenstate parity structure arise due to gap closing and reopening events between many-body Floquet bands whose photon sectors differ by an odd (even) number, the system admits corner MPMs (MZMs). In the system under our study, we however find that no MZMs are observed at all parameter values considered in our numerics. Therefore, in the rest of this paper, we will only focus on characterising the system's MPMs.

In addition to evaluating the relative eigenstate parity between $(0,0)$ and $(\pi,\pi)$ above, it is in general also necessary to inspect the relative eigenstate parity between the end points of the half anti-diagonal line $(0,0)$ and $(\pm \pi, \mp \pi)$. Together, these result in two $Z_2$ invariants labelled $\nu_{\rm d}$ and $\nu_{\rm ad}$ below. Under OBC, they signal the presence of MPMs at (respectively) two diagonal and anti-diagonal corners. However, as our analytical calculation shows in Sec.~\ref{appr} and Appendix~\ref{app1}, these invariants are given by the same expression $\nu_{\rm d}=\nu_{\rm ad}\equiv\nu_\pi$. This suggests that our system hosts either four MPMs (one at each corner) or none at all.   

}


\subsection{$Z_2$ invariant calculation}
\label{appr} 

To physically highlight the role of periodic drives designed above in generating nontrivial topology, we define and derive a bulk $Z_2$ invariant $\nu_\pi'$ by considering only the approximate $4\times 4$ truncated Floquet Hamiltonian $\mathcal{H}'_{BdG}$. Such an approximation is made by keeping only two photon sectors $a=0,1$ of the infinite matrix $\mathcal{H}_{\rm BdG}$. Physically, this corresponds to taking into account processes involving the emission and absorption of a single photon at a time. We leave the full derivation of the $Z_2$ invariant ($\nu_\pi$) based on the exact infinite matrix $\mathcal{H}_{\rm BdG}$ in Appendix~\ref{app1}, which is mathematically more involved but does not introduce new physics. For further simplifications, we also set $J_{s,x}=J_{s,y}=0$ throughout this section.

Under such a two-photon-sectors approximation, the infinite matrices $\xi_i$'s and $\eta_i$'s defined in Eqs.~(\ref{infp}) and (\ref{infp2}) reduce to the same set of $2\times 2$ Pauli matrices, which in the following will be denoted as $\tau_i$'s. We may then write $\mathcal{H}'_{\rm BdG}(\mathbf{k})$ along a diagonal and anti-diagonal line as
\begin{eqnarray}
\mathcal{H}'_{\rm BdG,d}(k)&=&\mathcal{H}'_{\rm BdG}(k_x=k_y=k) \;, \nonumber \\
&=& \frac{\hbar \omega}{2} \left(1 + \tau_z\right) +\mu \sigma_z + 2\sqrt{2}\Delta  \sin(k) \sigma_1 \nonumber \\
&&   + J_0 \cos(k) \sigma_z \tau_1  \;, \nonumber \\
\mathcal{H}'_{\rm BdG,ad}(k)&=&\mathcal{H}'_{\rm BdG}(k_x=-k_y=-k) \;, \nonumber \\
&=& \frac{\hbar \omega}{2} \left(1 + \tau_z\right) +\mu \sigma_z + 2\sqrt{2}\Delta \sin(k) \sigma_2 \nonumber \\
&&   + J_0 \cos(k) \sigma_z \tau_1  \;, 
\end{eqnarray}
where 
\begin{eqnarray}
\sigma_1 &=& \frac{1}{\sqrt{2}}(\sigma_x+\sigma_y)\;, \nonumber \\
\sigma_2 &=& \frac{1}{\sqrt{2}}(\sigma_x-\sigma_y)\;, \nonumber \\
\tau_1 &=& \frac{1}{J_0}(J_{0,x}\tau_x+J_{0,y}\tau_y)\;, \nonumber \\
J_0 &=& \sqrt{J_{0,x}^2+J_{0,y}^2} \;.
\end{eqnarray}

Ignoring the identity term, proper basis transformation allows us to rewrite $\mathcal{H}'_{\rm BdG,d}(k)$ (similarly for $\mathcal{H}'_{\rm BdG,ad}(k)$) in the block anti-diagonal form
\begin{eqnarray}
\mathcal{H}'_{\rm BdG,d}(k) &=& \left(\begin{array}{cc}
\mathbf{0} & W(k) \\
W^\dagger(k) & \mathbf{0} \\    
\end{array}\right) \;, \label{trun} 
\end{eqnarray}
where we have defined
\begin{eqnarray}
W(k) &=& \frac{\hbar \omega}{2} \tau_2 -\mathrm{i} \mu \tau_1 +2\sqrt{2}\Delta \sin(k)  - \mathrm{i} J_0 \cos(k) \;,  \nonumber \\
\end{eqnarray} 
with $\tau_2=-\mathrm{i}\tau_z\tau_1$. More explicitly, Eq.~(\ref{trun}) is obtained by applying the unitary transformation $\mathcal{H}'_{\rm BdG,d}(k) \rightarrow U \mathcal{H}'_{\rm BdG,d}(k) U^\dagger$ with
\begin{equation}
U=\exp\left(\mathrm{i}\frac{\pi}{4} \sigma_1 \tau_1\right), \label{unit}
\end{equation}
which brings $\mathcal{H}_{\rm BdG,d}'(k)$ to block anti-diagonal form in the $\sigma_z$ representation.

The Floquet eigenstate winding along (without loss of generality) the blue dashed line of Fig.~\ref{pic6} can then be defined and calculated as 
\begin{eqnarray}
n_{\rm d}' &=& \frac{1}{2\pi \mathrm{i}} \int_0^{\pi} \mathrm{Tr}\left[W^{-1}(k) \frac{d}{dk} W(k)\right] dk \nonumber \\
&=& \frac{1}{4\pi \mathrm{i}}\oint \left(\frac{w_+'(z)}{w_+(z)}+ \frac{w_-'(z)}{w_-(z)}\right) dz  \nonumber \\
&=& \begin{cases}
0 & \text{if } \left(\frac{\hbar^2 \omega^2}{4}-\mu^2-8\Delta^2\right)\times \left(\frac{\hbar^2 \omega^2}{4}-\mu^2+J_0^2\right)>0 \\
1 & \text{if } \left(\frac{\hbar^2 \omega^2}{4}-\mu^2-8\Delta^2\right)\times \left(\frac{\hbar^2 \omega^2}{4}-\mu^2+J_0^2\right)<0 \\
\end{cases} \;, \label{windapp} \nonumber \\
\end{eqnarray}
where 
\begin{eqnarray}
z &=& 2 \sqrt{2} \Delta \sin(k) - \mathrm{i} J_0 \cos(k) \;, \nonumber \\
w_+(z) &=& z+\sqrt{\frac{\hbar^2 \omega^2}{4}-\mu^2} \;, \nonumber \\
w_-(z) &=& z-\sqrt{\frac{\hbar^2 \omega^2}{4}-\mu^2}  \;, \label{compl}
\end{eqnarray}
and Cauchy residue theorem has been applied to obtain the last line. The same result is also obtained when a similar quantity is evaluated along one of the anti-diagonal lines, i.e., $n_{\rm ad}'=n_{\rm d}'$. 

Physically, the winding number calculated above counts the number of twists (gap closing and reopening) in the quasienergy bands associated with the Floquet BdG Hamiltonian along a half diagonal or anti-diagonal line. Although such bands do not represent the actual many-body quasienergy bands, they serve as the system's quasienergy excitations, i.e., quasienergies above a reference many-body Floquet band. It thus follows that a twist appearing in the quasienergy excitation spectrum directly translates to a twist in the full many-body spectrum. As a result, the above winding number may also faithfully count the number of twists in the many-body bands' eigenstate parity structure along a half diagonal or anti-diagonal line, thus representing the system's actual Floquet eigenstate winding. In the rest of this paper, quasienergy excitations will simply be referred to as quasienergies for simplicity, whereas the system's actual quasienergies are referred to as many-body quasienergies.

The presence or absence of Majorana modes is determined by the relative eigenstate parity between two end points of a half diagonal or anti-diagonal line \cite{Kit}. It can be obtained by taking the parity of the calculated Floquet eigenstate winding, which leads to the $Z_2$ invariants
\begin{eqnarray}
\nu_{\rm d}' &=& (-1)^{n_{\rm d}'} \nonumber \\
&=& \mathrm{sgn}\left(\frac{\hbar^2 \omega^2}{4}-\mu^2-8\Delta^2\right) \times \mathrm{sgn}\left(\frac{\hbar^2 \omega^2}{4}-\mu^2+J_0^2\right) \;, \nonumber \\
\nu_{\rm ad}' &=& (-1)^{n_{\rm ad}'} \nonumber \\
&=& \mathrm{sgn}\left(\frac{\hbar^2 \omega^2}{4}-\mu^2-8\Delta^2\right)\times \mathrm{sgn}\left(\frac{\hbar^2 \omega^2}{4}-\mu^2+J_0^2\right) \;. \nonumber \\ \label{appv}
\end{eqnarray} 
Since both expressions are identical, we can define a single $Z_2$ invariant $\nu_\pi'\equiv\nu_{\rm d}'=\nu_{\rm ad}'$, such that the system under consideration supports four MPMs at its corners or none at all whenever $\nu_\pi'=-1$ or $\nu_\pi'=1$ respectively. 

Note that the first (second) quantity on the right hand side of Eq.~(\ref{appv}) is always equal to $-1$ ($+1$) in the regime $\mu>\frac{\hbar \omega}{2}$ ($\mu<\frac{\hbar \omega}{2}$), i.e., $\nu_\pi'$ is independent of $\Delta$ ($J_0$). This allows us to compare $\nu_\pi'$ above with the exact invariant $\nu_\pi$ in the regimes $\mu>\frac{\hbar \omega}{2}$ and $\mu<\frac{\hbar \omega}{2}$ separately. In particular, at small parameter values $\mu$, $\Delta$ and $J_0$, we find that $\nu_\pi'$ coincides with the actual $\nu_\pi$, which in the regime $\mu<\frac{\hbar \omega}{2}$ is given as (see Appendix~\ref{app1} for technical detail), 
\begin{equation}
\nu_{\pi} = \prod_{n=0}^\infty \mathrm{sgn}\left[(2n+1)^2\frac{\hbar^2\omega^2}{4}-\mu^2 -8\Delta^2 \right]\;. \label{exac2}
\end{equation}
In general, however, $\nu_\pi'$ of Eq.~(\ref{appv}) does not the capture additional (e.g., nontrivial to trivial) transitions that occur at larger $\Delta$ values. For example, when
\begin{equation}
\frac{25\hbar^2\omega^2}{4}-\mu^2>8\Delta^2>\frac{9\hbar^2\omega^2}{4}-\mu^2 \;,
\end{equation} 
the actual $Z_2$ invariant $\nu_\pi=1$ predicts a topologically trivial regime with no corner MPMs, whereas $\nu_\pi'=-1$ continues to (incorrectly) predict a topologically nontrivial regime. In the regime $\mu>\frac{\hbar \omega}{2}$, the calculation of actual $\nu_\pi$ proves to be more cumbersome and we are unable to present its closed expression. However, while $\nu_\pi'$ predicts the emergence of corner MPMs at $J_0^2>\mu^2-\frac{\hbar^2 \omega^2}{4}$ when $\mu>\frac{\hbar \omega}{2}$, we find that corner MPMs are absent in this regime. There might still be another topological phase transition induced by $J_0$ in this case, but it occurs at a significantly different value of $J_0$ that is no longer well captured by $\nu_\pi'$.

The results presented so far show that the expected MPMs are truly of dynamical origin, whose existence can be traced back from the presence of nontrivial Floquet eigenstate winding induced by the topology of the time-periodic drives. That is, with the introduction of two time-periodic terms $h_c(t)\propto \cos(\omega t)$ and $h_s(t)\propto \sin(\omega t)$, the nontrivial winding number of $h_c(t)+\mathrm{i} h_s(t)$ with respect to time leads to the emergence of an additional set of anticommuting operators. Together with the existing $2\times 2$ Pauli matrices associated with particle-hole degree of freedom, they result in the possibility of properly defining and achieving nontrivial invariants. To further emphasize the importance of this aspect, we end this section by discussing the fate of the above invariants in the absence of any periodic drives and in the presence of topologically trivial drives.

In the absence of any periodic drives, the Floquet eigenstate winding $n_d'$ or $n_d$ and, consequently, $\nu_\pi'$ or $\nu_\pi$ are ill-defined. To support this statement, suppose we attempt to define $n_d'$ or $n_d$ by taking the limit of $J_{0,x}, J_{0,y}\rightarrow 0$ in Eqs.~(\ref{windapp}) or (\ref{exacwind}), while keeping $\frac{J_{0,y}}{J_{0,x}}$ constant to allow the infinite matrices $\eta_i$'s to remain being well-defined. In this case, however, we also have the freedom to consider an arbitrary value of the drives' frequency. By inspecting either Eq.~(\ref{windapp}) or (\ref{exacwind}), it then follows that depending on the frequency used to approach the limit, we may get either $n_d=1$, $n_d=0$, or even $n_d$ being undefined altogether (such as when $\hbar^2\omega^2/4-\mu^2-8\Delta^2=0$). This shows that the static limit of the above invariants do not exist, which is also consistent with the fact that a minimum of four bands is necessary to properly define a bulk invariant characterising a second-order topological phase in the spirit of Refs.~\cite{HTI1,HTI2}.

Related to the above argument, we should also emphasize that one may rule out the possibility of defining a static bulk invariant by instead evaluating the static limit of another set of topological invariants $\tilde{\nu}_{d}$ and $\tilde{\nu}_{ad}$ characterising the potential existence of corner MZMs in the driven setting. While we are not going to explicitly calculate such invariants in this paper, one may note that due to the $\tilde{\mathcal{P}}$ symmetry, it is possible to follow similar steps presented in Appendix~\ref{app1} to define two winding numbers $\tilde{n}_{\rm d}$ and $\tilde{n}_{\rm ad}$, whose parity corresponds to $\tilde{\nu}_{d}$ and $\tilde{\nu}_{ad}$. In particular, such winding numbers are obtained by evaluating contour integrations with respect to appropriately defined complex quantities that depend on $\Delta$, $J_{0,x}$, and $J_{0,y}$. Due to the absence of MZMs in the system, we expect that such contour integrations typically enclose an even number of poles at generic parameter values, thus leading to trivial $\tilde{\nu}_{d}$ and $\tilde{\nu}_{ad}$ values. However, at certain fine-tuned parameter values that depend on the system's frequency, there is also a possibility that some paths of such contour integrations intersect the poles. In such cases, $\tilde{\nu}_{d}$ and $\tilde{\nu}_{ad}$ consequently become ill-defined. By the same argument presented before, i.e., due to the freedom in choosing the driving frequency, the static limit of $\tilde{\nu}_{d}$ and $\tilde{\nu}_{ad}$ may therefore not exist, { as one may choose to evaluate the limit along a frequency value at which $\tilde{\nu}_{d}$ and $\tilde{\nu}_{ad}$ are ill-defined. It should be emphasized however that in the driven setting, where a fixed frequency value is considered, $\tilde{\nu}_{d}$ and $\tilde{\nu}_{ad}$ may still be well-defined. In this case, $\tilde{\nu}_{d}$ and $\tilde{\nu}_{ad}$ may still serve as valid topological invariants to characterize MZMs in the driven system.}

Finally, If topologically trivial drives are instead employed, e.g., with both $h_c(t),h_s(t)\propto \cos(\omega t)$, it may at first seem that the derivation presented above can be repeated to arrive at Eq.~(\ref{windapp}). In this case, however, the quantities $n_d'$ or $n_d$ and $\nu_\pi'$ or $\nu_\pi$ are no longer physically meaningful. This is because the symmetries $\mathcal{M}_{D}$ and $\mathcal{M}_{\rm AD}$ further map    
\begin{eqnarray}
\mathcal{M}_{\rm D} \tilde{\mathcal{H}}(k_x=\pi-k_y) \mathcal{M}_{\rm D}^{-1} &=& -\tilde{\mathcal{H}}(k_x=\pi-k_y)\;, \nonumber \\
\mathcal{M}_{\rm AD} \tilde{\mathcal{H}}(k_x=k_y-\pi) \mathcal{M}_{\rm AD}^{-1} &=& -\tilde{\mathcal{H}}(k_x=k_y-\pi)
\end{eqnarray}  
in addition to their action described in Eq.~(\ref{sym0}). { While the quantity $n_d'$ or $n_d$ may still remain being well-defined on its own, it no longer uniquely captures the second order topology of the whole 2D system. In particular, one may define another quantity $\tilde{n}_d'$ or $\tilde{n}_d$ that represents Floquet quasienergy winding along any curve, e.g., $\mathbf{k}=(k,\pi-k)$, respecting the same $\mathcal{M}_{\rm D}$. Since MPMs located at the system's two diagonal corners, if exist, are protected by $\mathcal{M}_{\rm D}$, $n_d'$ and $n_d$ must be equal if they were to represent a valid topological invariant. However, it follows that this may not always be the case. For example, in the case $J_{0,x}=J_{0,y}$ and $J_{s,x}=J_{s,y}$, $n_d'$ is still given by Eq.~(\ref{appv}), while $\tilde{n}_d'$ evaluated along $\mathbf{k}=(k,\pi-k)$ instead results in a trivial value $0$ at all parameter values. Similar argument holds with respect to the invariant $n_{\rm ad}'$ or $n_{\rm ad}$. This shows that similar to its static counterpart, $\nu_\pi$ is also ill-defined when the system is instead subjected to topologically trivial drives. Consequently, as we have also verified in Fig.~\ref{phase} of Appendix~\ref{app2}, no MPMs are expected to emerge in the system under such a driving protocol when OBC are introduced. }



\section{Numerical results}
\label{exat}

We will now verify numerically the predicted corner MPMs at parameter values for which $\nu_\pi=-1$. To this end, we directly construct the Floquet Hamiltonian associated with Eq.~(\ref{sys}), truncated up to a reasonably large maximum photon index $n_{\rm max}$ to allow numerical processing, then diagonalize it and accept only quasienergy solutions within $\left(0, \hbar \omega\right]$ \cite{note2}. Alternatively, such quasienergy solutions can also be obtained by diagonalizing the one-period time evolution operator (obtained numerically, e.g., via the use of split-operator method), which inherently takes into account all photon sectors in the Floquet Hamiltonian language. We have employed both approaches and obtained similar results. As such, unless otherwise specified, in the following we only present our results based on the former approach. 

\begin{figure}
	\begin{center}
		\includegraphics[scale=0.5]{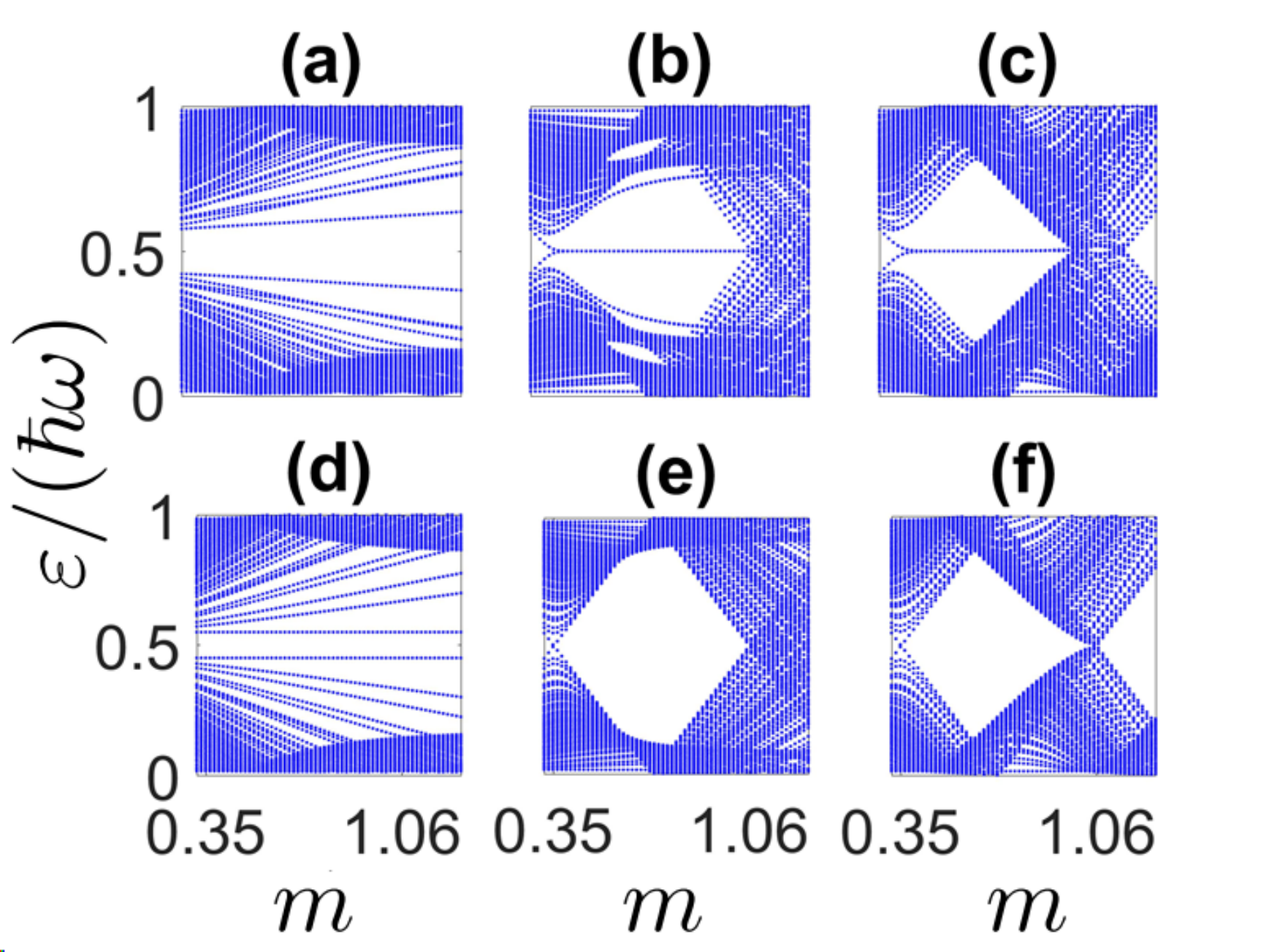}
		\caption{ Quasienergy spectrum of Eq.~(\ref{sys}) under (a,b,c) OBC and (d,e,f) PBC in both directions. In panels (a) and (d), only $J_{0,x}=J_{0,y}=m\hbar \omega$ is varied, while $\Delta= \frac{\hbar \omega}{2\pi}$ is fixed. In panels (b) and (e), only $\Delta =\frac{m}{2}\hbar \omega$ is varied, while $J_{0,x}=J_{0,y}= \frac{\hbar \omega}{\pi}$ is fixed. In panels (c) and (f), $J_{0,x}=J_{0,y}=2\Delta=m\hbar \omega$ is varied. In all panels, we set $J_{s,x}=J_{s,y}=0$, $\mu=\frac{0.1}{2\pi}\hbar \omega$, include up to $\pm 3$ photon sectors of the Floquet Hamiltonian (i.e., $n_{max}=3$) in our numerics, and take the system size to be $15\times 15$.}
		\label{pic1}
	\end{center}
\end{figure}

Figure~\ref{pic1} shows the calculated quasienergy solutions (under both PBC and OBC in both directions for side-by-side comparisons) as some system parameters are varied. There, we observe that corner MPMs (indicated by additional quasienergy solutions at $\frac{\hbar \omega}{2}$ in panels (b) and (c)) first emerge after two quasienergy bands touch at $\Delta_1= \sqrt{\frac{1}{8}-\frac{4\mu^2}{\hbar^2 \omega^2}} \hbar \omega \approx 0.35\hbar \omega$, where $\nu_\pi$ switches from $1$ to $-1$. Another quasienergy band touching occurs at $\Delta_2=\sqrt{\frac{9}{8}-\frac{4\mu^2}{\hbar^2 \omega^2}}\hbar \omega \approx 1.06\hbar \omega$, which switches $\nu_\pi$ from $-1$ back to $1$, followed by the absence of corner MPMs. Moreover, we note that varying $J_{0,x}$ and $J_{0,y}$ does not induce topological phase transition in the $\mu<\frac{\hbar \omega}{2}$ regime, which thus agrees with the analytical expression of $\nu_\pi$ presented in the previous section. On the other hand, by comparing panels (b,e) and (c,f) in Fig.~\ref{pic1}, it is evident that $J_{0,x}$ and $J_{0,y}$ may still affect the qualitatitve features of the observed quasienergy bands, especially at larger values of other system parameters (e.g., $\Delta$). In particular, the system may instead appear to become gapless at $\Delta>\Delta_2$ if $J_{0,x}$ and $J_{0,y}$ are fixed at small values. Physically, this can be understood as follows. Quasienergy gap closing events occurring at $\Delta>\Delta_1$ values are a result of higher-order photon emission and absorption processes. In this case, the mass terms capable of reopening the gap must consequently couple more than two adjacent photon sectors in the Floquet Hamiltonian. These can only be achieved by introducing either higher harmonic drives or large enough first harmonic driving strengths ($J_{0,x}$ and $J_{0,y}$).

Finally, we observe that at all parameter values considered in Fig.~\ref{pic1}, no gap is present around quasienergy zero, thus signifying the absence of MZMs. While not shown in the figure, we find that a gap around quasienergy zero might reopen at larger values of $\mu$, but no MZMs are observed in this case. As presented in the next section, however, chiral MMs around zero quasienergy may still emerge at some nonzero $J_{s,x}$ and $J_{s,y}$. This can be understood from the fact that the system under consideration may still host a first-order topological superconducting phase in the absence of periodic drives.

\begin{figure}
	\begin{center}
		\includegraphics[scale=0.5]{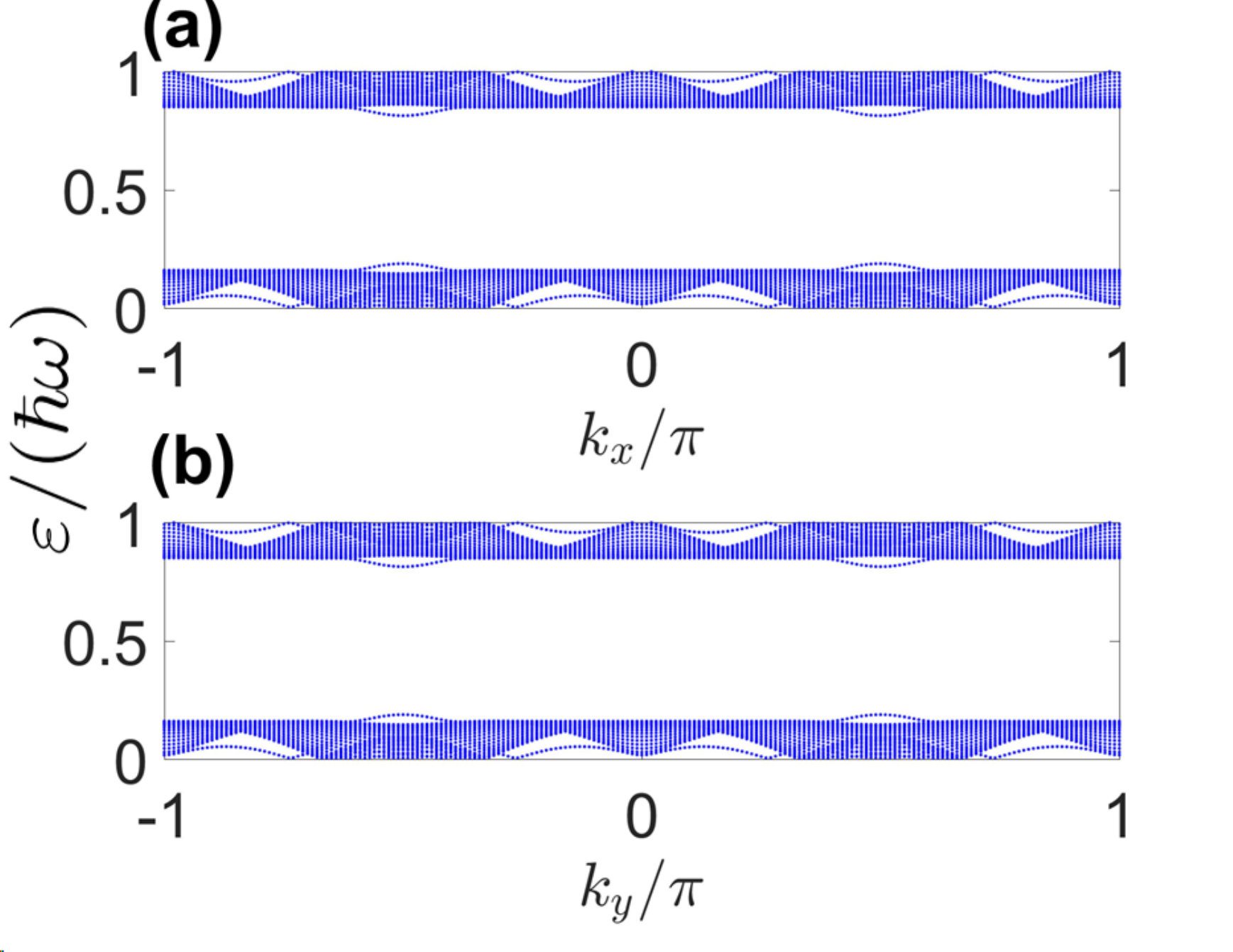}
		\caption{Quasienergy spectrum of Eq.~(\ref{sys}) under (a) OBC in the $y$-direction and PBC in the $x$-direction, (b) OBC in the $x$-direction and PBC in the $y$-direction. In both panels, $40$ sites are taken in the direction where OBC are applied and up to $\pm 3$ photon sectors of the Floquet Hamiltonian are included (i.e., $n_{max}=3$). The other parameters are set as $\mu=\frac{0.1}{2\pi}\hbar \omega$, $J_{0,x}=J_{0,y}=2\Delta = \frac{2}{\pi} \hbar \omega$, and $J_{s,x}=J_{s,y}=0$.}
		\label{pic2}
	\end{center}
\end{figure}

To further verify that MPMs observed in Fig.~\ref{pic1} are indeed corner and not edge modes, we plot in Fig.~\ref{pic2} the system's quasienergy spectrum under PBC in one direction and OBC in the other, which indeed shows the absence of $\frac{\hbar \omega}{2}$ solutions. In addition, we also explicitly calculate the support of each of the four observed quasienergy $\frac{\hbar \omega}{2}$ solutions (at a fixed set of parameter values for which $\nu_\pi=-1$) on Majorana operators representing the system's lattice sites. To this end, we first write each potential corner MPM as \cite{RG} 
\begin{equation}
\gamma_c(t)=\sum_{i,j,n} C_{i,j}^{(n)} \gamma_{i,j} \exp[\mathrm{i} (n-1/2)\omega t]\;, \label{Majsup1}
\end{equation}
where $\gamma_{i,j}$ is the Majorana operator defined in Eq.~(\ref{Majsup2}). { The coefficients $C_{i,j}^{(n)}$ can be determined from the real space Floquet BdG Hamiltonian as follows. In the Nambu-Sambe basis $(c_{i,j}^{(n)};c_{i,j}^{(n)\dagger})^T$, where $c_{i,j}^{(n)}$ is the Sambe vector representation of the fermionic operator $c_{i,j}e^{\mathrm{i}n\omega t}$, a quasienergy $\hbar \omega/2$ eigenvector of the real space Floquet BdG Hamiltonian can be written as $(w_{2i-1,j}^{(n)};w_{2i,j}^{(n)})^T$. Consequently, the operator 
\begin{equation}
\tilde{\gamma}_c^{(n)}=\sum_{i,j} \left(w_{2i-1,j}^{(n)}c_{i,j}^{(n)} +w_{2i,j}^{(n)}c_{i,j}^{(n)\dagger}\right) 
\end{equation}	
satisfies $\sum_m \left[\mathcal{H}_{n,m},\tilde{\gamma}_c^{(m)} \right]=\frac{\hbar \omega}{2}\tilde{\gamma}_c^{(n)}$, where $\mathcal{H}_{n,m}$ is defined in Eq.~(\ref{Sambe}). By Floquet theorem, it follows that $\gamma_c(t) =\sum_n \tilde{\gamma}_c^{(n)} \exp\left(\mathrm{i}(n-1/2)\omega t\right)$. Finally, by writing $c_{i,j}^{(n)}$ and $c_{i,j}^{(n)\dagger}$ in terms of $\gamma_{i,j}$ via Eq.~(\ref{Majsup2}), we obtain
\begin{eqnarray}
C_{2i-1,j}^{(n)} &=& w_{2i-1,j}^{(n)} +w_{2i,j}^{(n)} \;, \nonumber \\
C_{2i,j}^{(n)} &=& \mathrm{i} \left(w_{2i-1,j}^{(n)} -w_{2i,j}^{(n)}\right) \;. \label{relate} 
\end{eqnarray} 
}

\begin{figure}
	\begin{center}
		\includegraphics[scale=0.3]{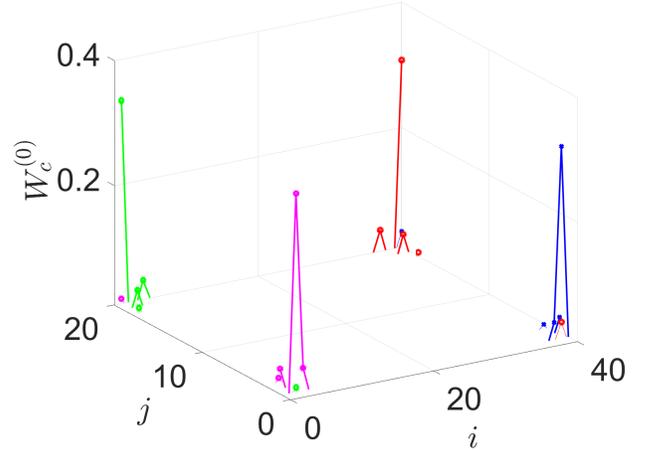}
		\caption{Support of each corner MPMs on Majorana operators representing the system's $20\times 20$ lattice sites (see Eq.~(\ref{weight})). Here, $i$ and $j$ represent the \emph{Majorana indices} in the $x$- and $y$-directions respectively (see Eq.~(\ref{Majsup2})). While the corner MPM solutions are obtained by numerically diagonalizing the truncated Floquet Hamiltonian containing up to $\pm 3$ photon sectors, only the zeroth photon sector contributions are shown. System parameters are chosen as $\Delta=\frac{\hbar \omega}{\pi}$, $\delta=\frac{0.1}{2\pi}\hbar \omega$, $J_{0,x}=\frac{3.4}{2\pi}\hbar \omega$, $J_{0,y}=\frac{3.6}{2\pi}\hbar \omega$, and $\mu=\frac{0.1}{2\pi}\hbar \omega$.}
		\label{pic3}
	\end{center}
\end{figure}

Given that the dominant contribution to $\gamma_c(t)$ comes from the zeroth photon sector, we plot in Fig.~\ref{pic3} the weights
\begin{equation}
[W^{(0)}_c]_{i,j}=|w_{i,j}^{(0)}|^2 \label{weight} 
\end{equation}
associated with the four quasienergy $\frac{\hbar \omega}{2}$ solutions in our system, where they are clearly localized at one of the four corners. There, we have also introduced a slight inhomogeneity of pairing strengths and hopping amplitudes in the $x$- and $y$-directions, i.e., $\Delta_x=\Delta +\delta$, $\Delta_y=\Delta -\delta$, and $J_{0,x}\neq J_{0,y}$, so as to demonstrate the robustness of such corner MPMs due to their topological nature. In Appendix~\ref{app2}, we further reveal that such corner MPMs are also robust against various other system imperfections   

\begin{figure}
	\begin{center}
		\includegraphics[scale=0.3]{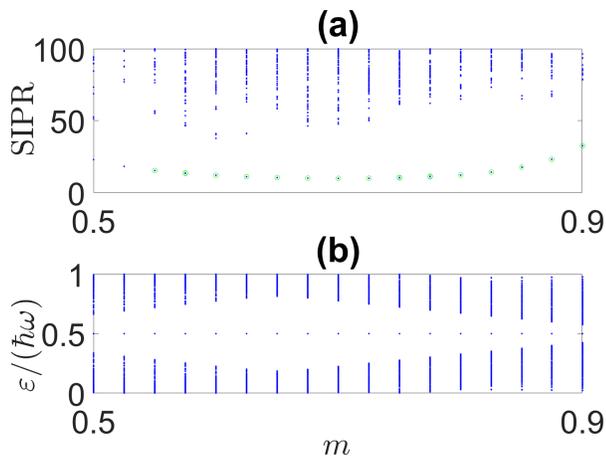}
		\caption{(a) SIPRs of all quasienergy eigenmodes as $J_{0,x}=J_{0,y}=2\Delta=m\hbar \omega$ is varied, while other system parameters are fixed at $J_{s,x}=J_{s,y}=0$, $\mu=\frac{0.1}{2\pi}\hbar \omega$ (those of MPMs are marked by green circles). (b) The associated quasienergy spectrum under the same parameter values as panel (a).}
		\label{IPR}
	\end{center}
\end{figure}

Finally, in order to quantitatively analyse the localization of the observed corner MPMs above, we define the stroboscopic inverse participation ratios (SIPRs) as follows. By first expanding a quasienergy eigenstate mode $\psi_\varepsilon(t)$ (i.e., an operator creating a quasienergy $\varepsilon$ from a reference state) in terms of Majorana operators defined in Eqs.~(\ref{Majsup1}) and (\ref{Majsup2}), i.e.,
\begin{equation}
\psi_\varepsilon(t)=\sum_{i,j,n} C_{\varepsilon,i,j}^{(n)} \gamma_{i,j}\exp\left(\mathrm{i} (n-1/2)\omega t\right)\;,
\end{equation}   
the { coefficients $C_{\varepsilon,i,j}^{(n)}$ are related to the quasienergy $\varepsilon$ eigenvector $(w_{\varepsilon,2i-1,j}^{(n)};w_{\varepsilon,2i,j}^{(n)})^T$ of the real space Floquet BdG Hamiltonian in the spirit of Eq.~(\ref{relate}).} The SIPR of $\psi_\varepsilon(t)$ is then given by
\begin{equation}
\mathrm{SIPR}[\psi_\varepsilon]=\frac{1}{\sum_{i,j} \left|\sum_n w_{\varepsilon,i,j}^{(n)}\right|^4} \;.
\end{equation}
Similar to its static counterpart, smaller SIPR signifies that a mode is more localized. In Fig.~\ref{IPR}(a), we plot the SIPRs of all the system's quasienergy eigenmodes in the regime where corner MPMs exist (see panel b for the associated quasienergy spectrum). There, we observe that the corner MPMs (marked by green circles), being localized near a system's corner, possess the lowest SIPRs and are clearly separated from those of other (bulk and/or edge) quasienergy eigenmodes. Moreover, by comparing both panels, SIPRs of these MPMs are observed to correlate with the system's quasienergy gap around $\varepsilon=\hbar\omega/2$ across different parameter values. This confirms the expected intuition that the localization length of corner MPMs scales inversely with such a quasienergy gap.   

\section{Discussions}

\label{coexist}

While the $Z_2$ invariant $\nu_\pi$ above was derived under the assumption that $J_{s,x}=J_{s,y}=0$ for simplicity, we have also verified that the observed corner MPMs remain robust at nonzero static hopping amplitudes $J_{s,x}$ and $J_{s,y}$. This is evidenced in Fig.~\ref{pic4}, where remarkably corner MPMs exist even at moderate values of $J_{s,x}$ and $J_{s,y}$. { Moreover, we also observe that chiral MMs additionally exist around zero quasienergy at some $J_{s,x}$ and $J_{s,y}$ values, as evidenced by the presence of quasienergy solutions in Fig.~\ref{pic4}(a) filling in the gap around zero quasienergy. Unlike the corner MPMs, which may only exist exclusively in the presence of periodic drives, these chiral MMs originate from the underlying static system under consideration (see Fig.~\ref{CMMpic}(a,c)), which corresponds to a first-order topologically nontrival superconductor in the regime $\mu<2(J_{s,x}+J_{s,y})$. It follows that the presence of periodic drives preserves such chiral MMs as long as the bulk gap around zero quasienergy remains open (see Fig.~\ref{CMMpic}(b,d))

The above discussion presents the possibility of an unprecedented scenario in which non-chiral and chiral MMs coexist in the same system. Such a feature is expected to find a promising application in quantum information processing, particularly for the task of quantum state transfers \cite{qst,qst2,qst3,qst4,qst5}. That is, one may consider the encoding of quantum information in some non-chiral MMs localized at corners of the one side of the system, transferring it to the chiral MMs, and retrieving it on the other side of the system by utilizing non-chiral MMs localized at its other corners. The detail and feasibility of this procedure will be left for future work.

\begin{figure}
	\begin{center}
		\includegraphics[scale=0.5]{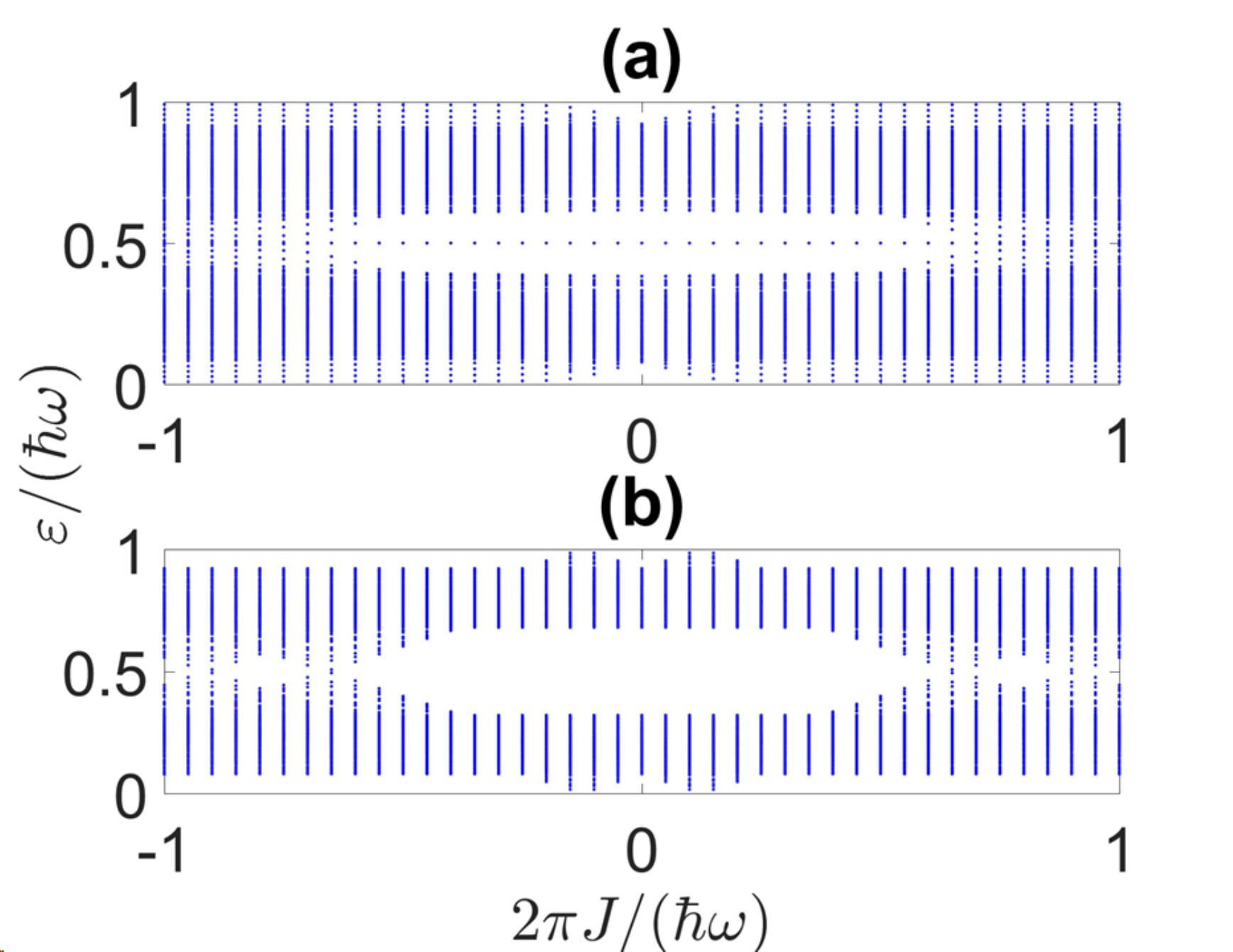}
		\caption{Quasienergy spectrum of Eq.~(\ref{sys}) under (a) OBC and (b) PBC in both directions as $J_{s,x}=J_{s,y}=J$ is varied. $15\times 15$ lattice sites are taken in panel (a), and the other parameter values are $J_{0,x}=J_{0,y}=\frac{\hbar\omega}{\pi}$, $\Delta=\frac{1.5}{2\pi}\hbar \omega$, and $\mu=\frac{1}{4\pi}\hbar \omega$ in both panels.}
		\label{pic4}
	\end{center}
\end{figure}

\begin{figure}
	\begin{center}
		\includegraphics[scale=0.5]{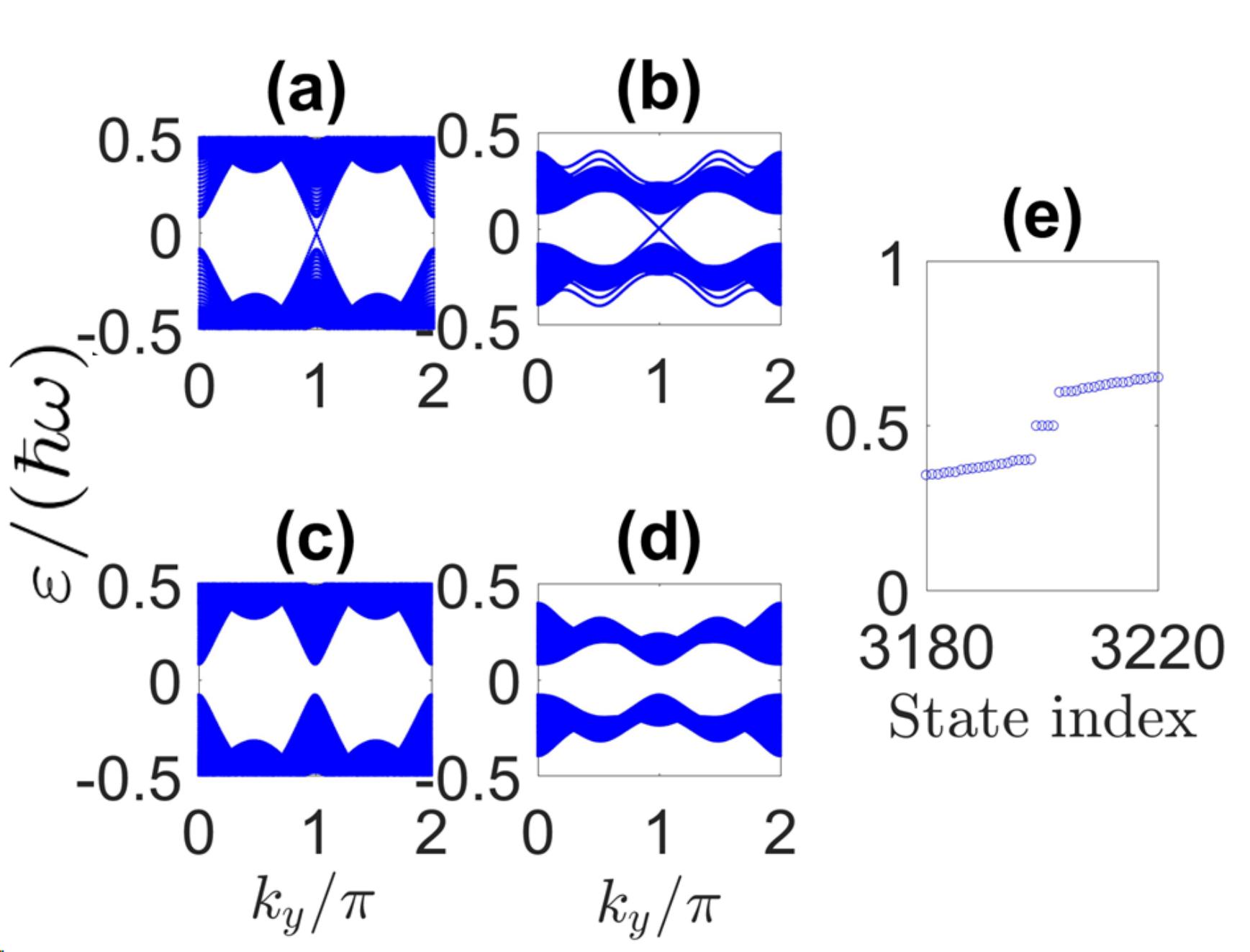}
		\caption{Quasienergy spectrum of Eq.~(\ref{sys}) under (a,b) OBC in the $x$-direction with 55 sites and PBC in the $y$-direction, (c,d) PBC in both directions, and (e) OBC in both directions with $15\times 15$ sites. Periodic driving parameters are set to (a,c) $J_{0,x}=J_{0,y}=0$ (static limit) and (b,d,e) $J_{0,x}=J_{0,y}=\frac{\hbar\omega}{\pi}$. The other parameter values are $\Delta=\frac{1.5}{2\pi}\hbar \omega$, and $\mu=J_{s,x}=J_{s,y}=\frac{1}{4\pi}\hbar \omega$ in all panels.}
		\label{CMMpic}
	\end{center}
\end{figure} 

Before ending this section, it is necessary to compare this work with relevant earlier literature on Floquet topological phases \cite{Rud,FHTI1,FHTI2,FHTI3,FHTI4,FHTI5,FHTI6,FHTI7,YP1,YP2,YP3}. { First, Ref.~\cite{Rud} demonstrates the possibility of generating nontrivial two-band Floquet time-reversal invariant topological insulators (TRIs), whose static counterparts also require a minimum of four bands. In such a construction, appropriate choice of periodic drives turns an inherently trivial system into a first-order topological one. By contrast, in the present work, we instead demonstrate the possibility of designing periodic drives that yield second-order topological systems from an otherwise inherently trivial system. In this case, it is expected that adapting our driving protocol to the model of Ref.~\cite{Rud} yields a second-order Floquet TRI, whereas the application of the driving scheme proposed in Ref.~\cite{Rud} to our model instead leads to a first-order Floquet topological superconductor. This highlights the main difference between the two works.}

Second, some of Refs.~\cite{FHTI1,FHTI2,FHTI3,FHTI4,FHTI5,FHTI6,FHTI7,YP1,YP2,YP3} demonstrate the generation of higher-order topologically nontrivial phases by applying appropriate time-periodic drives to a static topologically trivial system. However, the latter may already possess the necessary requirements to host such higher-order topological phases on its own, accomplished such as by either tuning some system parameters or adding appropriate mass terms. In this case, the time-periodic drives simply play the role of either system parameters renormalization or mass terms simulation, whose topology may thus (in principle) be traced back from the underlying static system. By contrast, the emergence of Floquet SOTSC in our system is only possible via the implementation of nontrivial topology (winding number) in the time-domain. In this case, the underlying static system may not even exhibit second-order topological characterization. 

To further elaborate the above argument, we shall compare our construction with that of Refs.~\cite{YP1,YP2,YP3}, which at first glance might look similar to ours (i.e., due to the use of monochromatic time-periodic drives). In Refs.~\cite{YP1,YP2,YP3}, the time-periodic drives are designed such that the resulting Hamiltonian obeys a time-glide symmetry, which can then be viewed as an effective reflection symmetry in the enlarged Hilbert (Sambe) space. In this case, the role of the time-periodic drives is to effectively create a symmetry necessary for the formation of second-order topological phases, whereas the underlying static Hamiltonian already contains the necessary topological structure. This is further evidenced by the fact that four-band models are used in these works, i.e., the minimum number of bands expected for the formation of second-order topology in static systems. By contrast, the static system considered in this paper corresponds to a two-band (first-order) chiral topological superconductor. The latter is incapable of exhibiting nontrivial second-order topology under any circumstances due to the lack of mass terms (with only one set of Pauli matrices available) to open the edge states' gap. On the other hand, the symmetries $\mathcal{P}$, $\tilde{\mathcal{P}}$, $\mathcal{M}_{\rm D}$, and $\mathcal{M}_{\rm AD}$ are already present, now described by the static operators $ \mathcal{P}=\tilde{\mathcal{P}}=\sigma_x \mathcal{K}$, $\mathcal{M}_{\rm D}=\frac{1}{\sqrt{2}} (\sigma_x-\sigma_y)$ and $\mathcal{M}_{\rm AD}=\frac{1}{\sqrt{2}} (\sigma_x+\sigma_y)$. In this case, the periodic drives genuinely facilitate the emergence of additional winding invariant in the system, thus enabling $\nu_\pi$ to be properly defined and take a nontrivial value.


\section{Concluding remarks}
\label{conc}

In this paper, we proposed the construction of Floquet SOTSC without internal (pseudo-spin or orbital) degrees of freedom or spatially modulating any system parameters. In this case, the interplay between topological superconductivity and nontrivial winding of the periodic drives in the time-domain provides the necessary ingredient for the emergence of truly dynamical Majorana modes at the system's corners with no static analogues. While we considered only a single set of periodic drives to demonstrate the physics at work, such time-induced topology can also be achieved for a class of other periodic drives (see e.g., those considered in Appendix~\ref{ddf}).

Following the above findings, various directions for potential future studies can be envisioned. In the area of Floquet engineering, alternative realizations of existing (first- or higher-)order topological phases with significantly simpler systems may be possible through the application of several appropriate time-periodic potentials exhibiting nontrivial winding number in the time domain. In the area of quantum computing, the relatively less demanding system's complexity for hosting time-induced MMs may offer a fresh perspective towards the physical realizations of large-scale Majorana qubit architectures. Moreover, the possibility of time-induced topological superconductors to host chiral and non-chiral MMs simultaneously may allow the design of Majorana-based quantum state transfer schemes, as briefly commented in Sec.~\ref{coexist}. Finally, we expect that the idea of time-induced topology may open up opportunities for the discovery of novel Floquet topological phases.  

\begin{acknowledgements}
	{\bf Acknowledgement}: This work is supported by the Australian Research Council Centre of Excellence for Engineered Quantum Systems (EQUS, CE170100009). The author thanks Longwen Zhou for carefully reading the first draft of this manuscript and providing useful comments. { The author thanks an anonymous referee for pointing out the existence of symmetry $\mathcal{P}'$ that allows the characterization of MZMs.}
\end{acknowledgements}

\appendix

\section{General derivation of $Z_2$ invariant $\nu_\pi$} \label{app1}

Due to the similarity between the algebra of $\eta_i$'s and $\xi_i$'s with that of $2\times 2$ Pauli matrices, the idea presented in Sec.~\ref{appr} can be readily generalized to obtain the actual $Z_2$ invariant associated with the infinite-dimensional Floquet Hamiltonian $\mathcal{H}_{\rm BdG}$. To this end, by continuing to take $J_{s,x}=J_{s,y}=0$ throughout this section for simplicity, we first apply a similar basis transformation that anti-diagonalizes $\mathcal{H}_{\rm BdG}$ via the unitary operator
\begin{equation}
U=\exp\left(\mathrm{i}\frac{\pi}{4} \sigma_1 \eta_1\right)\;, \label{unit2} 
\end{equation}
where $\eta_1\equiv\eta_x(\phi)$ and $\tan\phi=J_{0,y}/J_{0,x}$. This leads to a matrix in Eq.~(\ref{trun}) of Sec.~\ref{appr}, but with $W(k)$ now replaced by an infinite matrix $\mathcal{W}(k)$ of the form
\begin{eqnarray}
\mathcal{W}(k) &=& -\mathrm{i} \frac{\hbar \omega}{2} \xi_z\eta_1 -\mathrm{i} \mu \eta_1 +2\sqrt{2}\Delta \sin(k) \xi_0 - \mathrm{i} J_0 \cos(k) \xi_1 \eta_1\;.  \nonumber \\ \label{wmat}
\end{eqnarray}
The Floquet eigenstate winding can similarly be obtained by evaluating
\begin{eqnarray}
n_{\rm d} &=& \frac{1}{2\pi \mathrm{i}} \int_0^\pi  \mathrm{Tr}\left[\mathcal{W}^{-1}(k) \frac{d}{dk} \mathcal{W}(k)\right] dk \nonumber \\ 
&=& \sum_{n=0}^\infty \sum_{s=\pm}\frac{1}{2\pi \mathrm{i}} \int_0^\pi  w_{n,s}^{-1}(k) \frac{d}{dk} w_{n,s}(k) dk  \;, \label{genwin}
\end{eqnarray}
where $w_{n,\pm}$ are the eigenvalues of $\mathcal{W}(k)$. They can be obtained exactly when $J_0=0$ by explicitly writing down the infinite matrix of $\mathcal{W}(k)$ (ignoring the identity term $\xi_0$ for a moment),
\begin{widetext}
\begin{eqnarray}
\mathcal{W}(k) &=& \left(\begin{array}{cccccc}
\ddots & \vdots &\vdots & \vdots & \vdots & \iddots \\ 
\ldots & \mathbf{0} & \mathbf{0} & \mathbf{0} &  \color{blue} \left(-\mathrm{i}\frac{3\hbar \omega}{2}  -\mathrm{i} \mu\right) \exp\left(\mathrm{i} 3\phi\right) & \ldots \\
\ldots & \mathbf{0} & \mathbf{0} & \color{red} \left(-\mathrm{i}\frac{\hbar \omega}{2}  -\mathrm{i} \mu\right)\exp\left(\mathrm{i} \phi\right) & \mathbf{0} & \ldots \\
\ldots & \mathbf{0} & \color{red} \left(\mathrm{i}\frac{\hbar \omega}{2}  -\mathrm{i} \mu\right) \exp\left(-\mathrm{i} \phi\right) & \mathbf{0} & \mathbf{0} & \ldots \\
\ldots &  \color{blue} \left( \mathrm{i}\frac{3\hbar \omega}{2}  -\mathrm{i} \mu \right) \exp\left(-\mathrm{i} 3\phi\right) & \mathbf{0} & \mathbf{0} & \mathbf{0} & \ldots \\
\iddots & \vdots &\vdots & \vdots & \vdots & \ddots \\ 
\end{array}\right) \;. \nonumber \\ \label{matr}
\end{eqnarray}
\end{widetext}
In particular, a pair of photon sectors $n$ and $(1-n)$ (such as those marked by the same colour in Eq.~(\ref{matr})) is decoupled from the rest and can be individually diagonalized. Its eigenvalues can thus be labeled by the quantum numbers $n=0,1,\cdots$ (associated with different pairs of photon sectors) and $s=\pm 1$ (associated with the two eigenvalues within a given pair of photon sectors), which are given by (now taking into account the $\xi_0$ term as well),
\begin{equation}
w_{n,s}^{(0)}=2\sqrt{2} \Delta \sin(k) +s \sqrt{\frac{(2n+1)^2\hbar^2\omega^2}{4}-\mu^2} \;.
\end{equation} 
The full eigenvalues $w_{n,s}$ of $\mathcal{W}(k)$ when $J_0\neq 0$ can then be obtained perturbatively. To this end, we first write, 
\begin{eqnarray}
w_{n,s} &=& w_{n,s}^{(0)} + \sum_j A_{n,s,j} ( -\mathrm{i} J_0\cos(k))^j  \;,
\end{eqnarray}
where $A_{n,s,j}$ is a constant that generally depends on $\omega$ and $\mu$. Next, we note that the (right) eigenvector $|w_{n,s}^{(0)}\rangle$ associated with $w_{n,s}^{(0)}$ has only two nonzero elements at row $n$ and $1-n$, i.e., $|w_{n,s}^{(0)}\rangle_n = 1$ and $|w_{n,s}^{(0)}\rangle_{1-n} = s\exp\left(\chi_n -\mathrm{i} \phi\right)$ respectively where $\tanh \chi_n = \frac{(2n+1)\hbar \omega}{2\mu}$. On the other hand, any power of the infinite matrix $\xi_1 \eta_1$ connects only photon sectors of the same parity. By noting that shifts in $w_{n,s}^{(0)}$ are obtained by evaluating terms of the form
\begin{equation}
\frac{\langle w_{n,s} | \xi_1 \eta_1 |w_{n_1,s_1}\rangle \langle w_{n_1,s_1} | \xi_1 \eta_1 |w_{n_2,s_2}\rangle \cdots \langle w_{n_j,s_j} | \xi_1 \eta_1 |w_{n,s}\rangle   }{(w_{n,s}^{(0)}-w_{n_1,s_1}^{(0)})(w_{n,s}^{(0)}-w_{n_2,s_2}^{(0)})\cdots (w_{n,s}^{(0)}-w_{n_j,s_j}^{(0)})}\;, \label{perterm}
\end{equation}
and further realizing that
\begin{widetext}
\begin{eqnarray}
\sum_{s_1,\cdots,s_j=\pm} \frac{\langle w_{n,+} | \xi_1 \eta_1 |w_{n_1,s_1}\rangle \langle w_{n_1,s_1} | \xi_1 \eta_1 |w_{n_2,s_2}\rangle \cdots \langle w_{n_j,s_j} | \xi_1 \eta_1 |w_{n,+}\rangle}{(w_{n,+}^{(0)}-w_{n_1,s_1}^{(0)})(w_{n,+}^{(0)}-w_{n_2,s_2}^{(0)})\cdots (w_{n,+}^{(0)}-w_{n_j,s_j}^{(0)})}&=& \nonumber \\
(-1)^j\sum_{s_1,\cdots,s_{j}=\pm} \frac{\langle w_{n,-} | \xi_1 \eta_1 |w_{n_1,s_1}\rangle \langle w_{n_1,s_1} | \xi_1 \eta_1 |w_{n_2,s_2}\rangle \cdots \langle w_{n_j,s_j} | \xi_1 \eta_1 |w_{n,-}\rangle}{(w_{n,-}^{(0)}-w_{n_1,s_1}^{(0)})(w_{n,-}^{(0)}-w_{n_2,s_2}^{(0)})\cdots (w_{n,-}^{(0)}-w_{n_j,s_j}^{(0)})} && \;,
\end{eqnarray}
\end{widetext}
where $n_1,\cdots, n_j \neq n$, it follows that $A_{n,s,2j}=s A_{n,2j}$, $A_{n,s,2j-1}=A_{n,2j-1}$, and $A_{n,s,2j-1}=0$ for $j\leq n$. We then arrive at
\begin{eqnarray}
w_{n,s} &=& w_{n,s}^{(0)} + \sum_{j=1}^{\infty} (-1)^j s A_{n,2j} \left[J_0\cos(k)\right]^{2j} \nonumber \\
&& +\mathrm{i} \; \sum_{j=n+1}^\infty (-1)^j A_{n,2j-1} \left[J_0\cos(k)\right]^{2j-1}\;. 
\end{eqnarray} 

We may now define
\begin{equation}
z_n = 2\sqrt{2} \Delta \sin(k) +\mathrm{i} \sum_{j=n+1}^\infty (-1)^j A_{n,2j-1} \left[J_0\cos(k)\right]^{2j-1}
\end{equation}
and turn Eq.~(\ref{genwin}) into a contour integration
\begin{equation}
n_{\rm d} = \sum_{n=0}^\infty \sum_{s=\pm}\frac{1}{4\pi \mathrm{i}} \oint  \frac{w_{n,s}'(z_n)}{w_{n,s}(z_n)} dz_n \;. \label{contour}
\end{equation}
Let us first assume that $\mu<\frac{\hbar \omega}{2}$. By applying residue theorem, we may identify poles along the real axis at $z_n=\mathcal{Z}_n=\pm \sqrt{\frac{(2n+1)^2\hbar^2 \omega^2}{4}-\mu^2}$ which lead to 
\begin{equation}
n_{\rm d} (\mu< \frac{\hbar \omega}{2}) =\sum_{n=0}^\infty \left(1- \mathrm{sgn}\left[(2n+1)^2\frac{\hbar^2\omega^2}{4}-\mu^2 -8\Delta^2 \right]\right)\;, \label{exacwind}
\end{equation}
or equivalently
\begin{equation}
\nu_{\rm d} (\mu< \frac{\hbar \omega}{2}) = \prod_{n=0}^\infty \mathrm{sgn}\left[(2n+1)^2\frac{\hbar^2\omega^2}{4}-\mu^2 -8\Delta^2 \right]\;. \label{exac}
\end{equation}
This generalizes the first quantity on the right hand side of Eq.~(\ref{appv}). The second quantity on the right hand side of Eq.~(\ref{appv}) can in principle be similarly generalized by considering $\mu>\frac{\hbar \omega}{2}$. In this case, contour integration of Eq.~(\ref{contour}) contains poles along the imaginary axis, which can be captured by varying $J_0$. However, the exact locations of these poles are also determined by the actual values of $A_{n,j}$. The latter can be obtained by explicitly evaluating many terms of the form Eq.~(\ref{perterm}). Such a calculation is very cumbersome and will thus not be pursued further here. Finally, we note that the analysis above can be repeated to obtain an identical expression for $\nu_{\rm ad}$. This allows us to define a single $Z_2$ invariant $\nu_\pi=\nu_{\rm ad}= \nu_{\rm d}$ presented in Eq.~(\ref{exac2}).


\section{Robustness of corner MPMs against realistic effects}
\label{app2} 

In the main text, we have assumed for simplicity that the system under consideration is ideal, i.e., it is free from disorders and has a perfect time-periodicity, as well as a perfect square-shaped geometry. In the following, we highlight the robustness of the system's corner MPMs when these assumptions are relaxed. 

\subsection{Spatial disorders}

We first consider the presence of spatial disorders on all system parameters 
\begin{eqnarray}
J_{0,x}&\rightarrow& J_{0,x}+\delta J_{0,x}^{(i,j)} \;, \nonumber \\
J_{0,y}&\rightarrow& J_{0,y}+\delta J_{0,y}^{(i,j)} \;, \nonumber \\
\Delta_x &=& \Delta +\delta \Delta_x^{(i,j)} \;, \nonumber \\
\Delta_y &=& \Delta +\delta \Delta_y^{(i,j)} \;, \nonumber \\
J_{s,x}&\rightarrow& J_{s,x}+\delta J_{s,x}^{(i,j)} \;, \nonumber \\
J_{s,y}&\rightarrow& J_{s,y}+\delta J_{s,y}^{(i,j)}\;, \label{disspat} 
\end{eqnarray}
where $\Delta_x$ and $\Delta_y$ are the pairing strengths in the $x$- and $y$-direction respectively. Values of the disorder parameters $\delta S^{(i,j)}$, where $S\in\left\lbrace J_{0,x}, J_{0,y}, \Delta_{x}, \Delta_{y}, J_{s,x}, J_{s,y}\right\rbrace$, are uniformly drawn from $\left[-\delta S, \delta S\right]$. The disorder averaged quasienergy levels of the system in  the vicinity of $\varepsilon=\frac{\hbar \omega}{2}$, under OBC in both directions, are arranged and depicted in Fig.~\ref{disd}(a), where four corner MPMs are still clearly observed. 


\subsection{Temporal noise}

We next consider the effect of temporal noise by evaluating the time-evolution of a corner MPM $\gamma_c(t)$ for $10$ periods, where the system parameters may slightly change after each period. To this end, we may again model all system parameters according to Eq.~(\ref{disspat}), where $\delta S^{(i,j)}\rightarrow \delta S^{(s)}$ for $(s-1)T<t<sT$ and each $\delta S^{(s)}$ is again uniformly drawn from $\left[-\delta S, \delta S\right]$. It is noted that with proper scaling of these system parameters, such a noise model also captures the effects of driving with imperfect periodicity. By writing $\gamma_c(t)$ in terms of Majorana operators as in Eqs.~(\ref{Majsup1}) and (\ref{relate}), we plot the weights (see Eq.~(\ref{weight})) of these Majorana operators supporting $\gamma_c(t)$ at $t=10T$ in Fig.~\ref{disd}(b) and (c). It is evident that the time-evolved MPM under such imperfect driving (panel c) remains localized near a corner and is qualitative similar to that in the ideal case (panel b). This demonstrates the robustness of the system's corner MPMs against temporal noise.


\begin{figure}
	\begin{center}
		\includegraphics[scale=0.33]{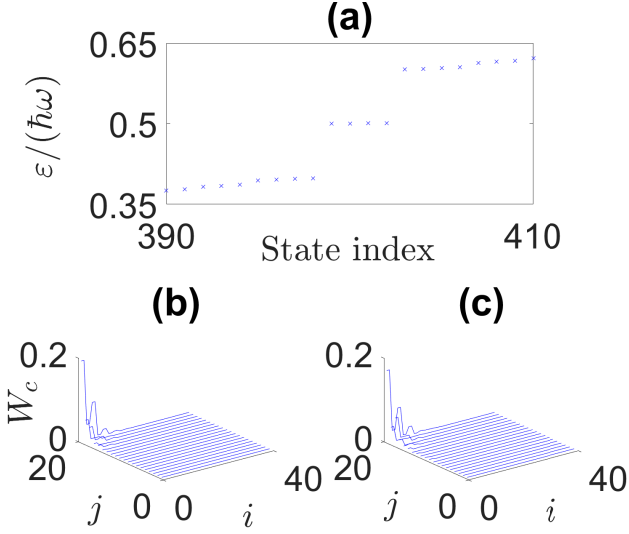}
		\caption{  (a) Quasienergy levels of the system under OBC in both directions with $20\times 20$ sites under disorder parameters $\delta J_{0,x}=\delta J_{0,y}=\delta\mu = 10 \delta \Delta_x= 10\delta \Delta_y =10\delta J_{s,x} = 10\delta J_{s,y}=\frac{\hbar \omega}{20 \pi}$, averaged over $50$ disorder realizations. (b,c) Support of the time-evolved $\psi_c(t=10T)$ on Majorana operators $\gamma_{i,j}$ representing the system's $20\times 20$ sites (see Eq.~(\ref{weight})) in the (b) absence and (c) presence of temporal disorders with $2\delta J_{s,x}=2\delta J_{s,y}=\delta J_{0,x}=\delta J_{0,y}=2\delta \Delta_x=2\delta \Delta_y=2\delta\mu=\frac{\hbar \omega}{20\pi}$. All other system parameters are set to $J_{0,x}=J_{0,y}=\frac{\hbar \omega}{\pi}$, $\Delta=\frac{1.5\hbar \omega}{2\pi}$, and $\mu=J_{s,x}=J_{s,y}=\frac{\hbar \omega}{4\pi}$. }
		\label{disd}
	\end{center}
\end{figure} 

\subsection{Geometric imperfections}

To simulate geometric imperfections, we introduce a defect near a system's corner by switching on a large value of chemical potential in the affected region. In Fig.~\ref{defect}, we observe that the MPM originally located at the bottom left corner in the ideal case remains well localized in the presence of defects with different sizes. This further demonstrates the robustness of such corner MPMs away from a perfect square-shaped system's geometry.       

\begin{figure}
	\begin{center}
		\includegraphics[scale=0.33]{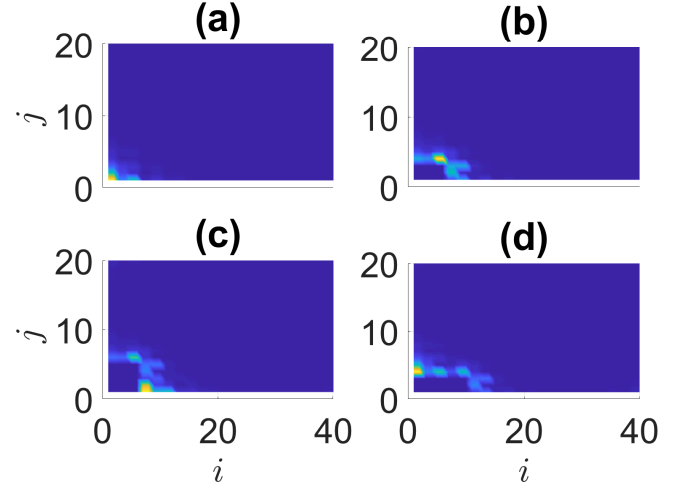}
		\caption{ Support of a corner MPM (higher values correspond to brighter colours) on Majorana operators representing the system's $20\times 20$ sites in the spirit of Eq.~(\ref{weight}) under the presence of (a) no defect, (b) square-shaped defect of size $3\times 3$ sites, (c) rectangle-shaped defect of size $3\times 5$ sites, (d) rectangle-shaped defect of size $5\times 3$ sites. System parameters are set to $J_{0,x}=J_{0,y}=\frac{\hbar \omega}{\pi}$, $\Delta=\frac{1.5\hbar \omega}{2\pi}$, and $\mu=J_{s,x}=J_{s,y}=\frac{\hbar \omega}{4\pi}$.}
		\label{defect}
	\end{center}
\end{figure} 

\subsection{Relative phase imperfection}

Another possible imperfection we may take into account concerns the deviation in the relative phase between the two drives of the system from $\pi/2$. That is, by now writing $J_x=J_{s,x}+J_{0,x}\cos(\omega t)$ and $J_y(t)=J_{s,y}+J_{0,y} \sin(\omega t + \xi)$ in Eq.~(\ref{sys}), we investigate the fate of the system's topology with respect to choosing $\xi\neq 0$. As Fig.~\ref{phase} shows, corner MPMs in fact also exist for any $\xi \neq \pi/2$, which can be understood as follows. Choosing a different value of $\xi$ amounts to modifying the value of $\phi$ appearing in $\mathcal{M}_{\rm D}$ and $\mathcal{M}_{\rm AD}$ of Eq.~(\ref{sym}) to $\phi=\arctan\left(\frac{J_{0,y}\cos(\xi)}{J_{0,x}+J_{0,y}\sin(\xi)}\right)$. In this case, the $Z_2$ invariant derivation presented in Sec.~\ref{appr} and Appendix~\ref{app1} proceeds in almost exactly the same way (the only difference being the form of unitary transformation used in bringing $\mathcal{H}_{\rm BdG,d}(k)$ to the anti-diagonal form), thus giving rise to the same $\nu_\pi$ expression (up to a redefinition of the quantity $J_0$). 

An exception to the above argument arises in the special case $\xi=\pi/2$, which leads to topologically trivial drives discussed in Sec.~\ref{appr}. That is, due to additional symmetrical lines at $\mathbf{k}=(k,\pi-k)$ and $\mathbf{k}=(k,k-\pi)$ with respect to $\mathcal{M}_{\rm D}$ and $\mathcal{M}_{\rm AD}$ respectively, the $Z_2$ invariant $\nu_\pi$ no longer represents a meaningful quantity. In this case, second-order topological characterization breaks down, and corner MPMs are not expected to be present.

\begin{figure}
	\begin{center}
		\includegraphics[scale=0.5]{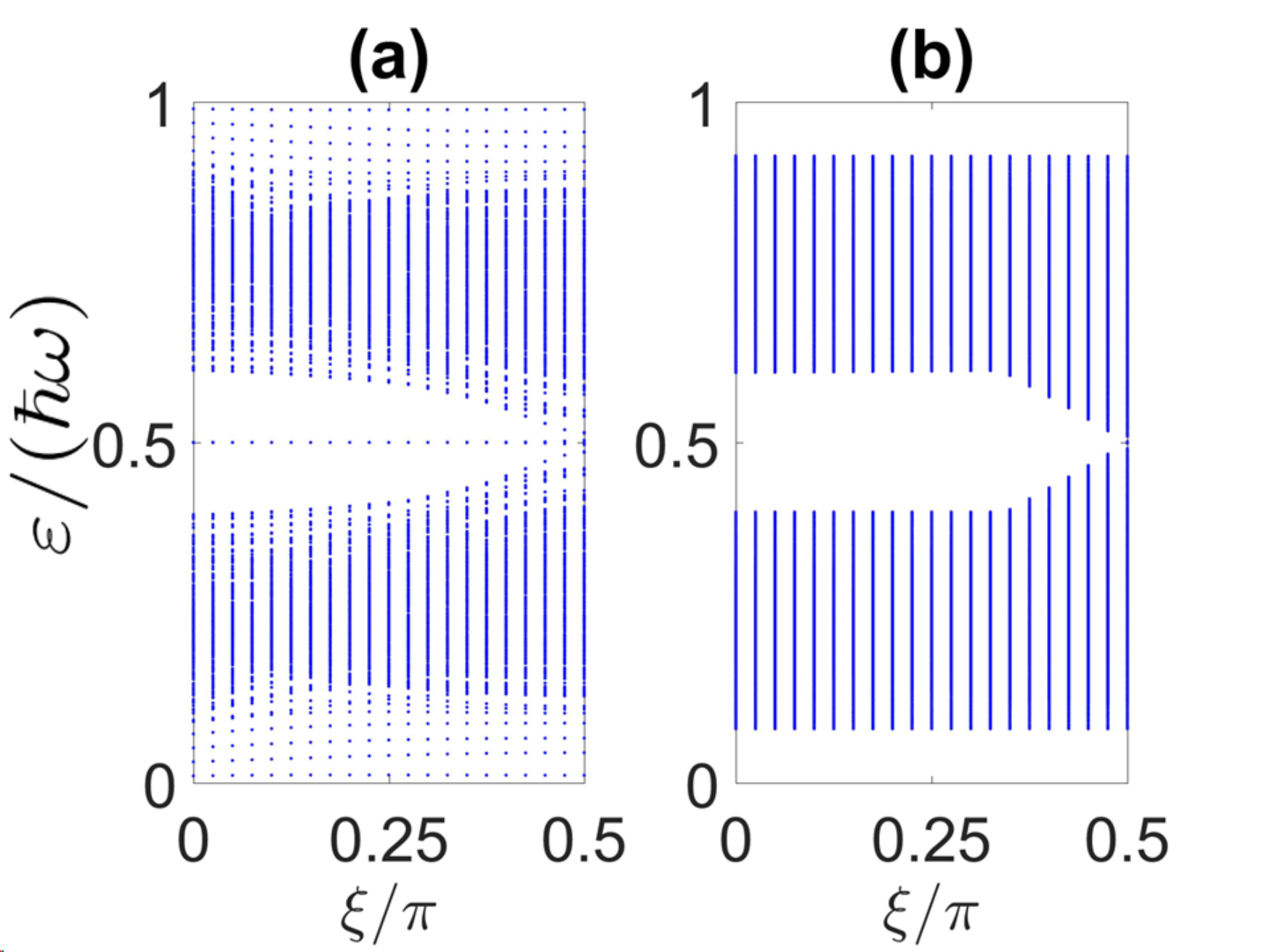}
		\caption{Quasienergy spectrum of Eq.~(\ref{sys}) as a function of the relative phase $\xi$ between $J_x(t)$ and $J_y(t)$ driving under (a) OBC and (b) PBC in both directions. All system parameters are set to $J_{0,x}=J_{0,y}=\frac{\hbar \omega}{\pi}$, $\Delta=\frac{1.5\hbar \omega}{2\pi}$, and $\mu=J_{s,x}=J_{s,y}=\frac{\hbar \omega}{4\pi}$. }
		\label{phase}
	\end{center}
\end{figure} 

\subsection{Heating effect}

Finally, another possible limitation of realizing Floquet closed systems in general concerns the effect of heating. That is, in the presence of particle-particle interactions, any generic initial state in such systems is hypothesized to eventually thermalize to a topologically trivial infinite temperature state. Aspects of heating in Floquet systems have been the subject of several studies in recent years, which involve a variety of different approaches \cite{heating1,heating10,heating2,heating3,heating4,heating5,heating6,heating7,heating8,heating9}. In the context of time-induced SOTSC introduced in this paper, properly analysing the effect of heating requires a more rigorous modelling of interactions that may be present in the system under consideration. As such, it deserves a separate study on its own and is beyond the scope of this paper. It is to be emphasized however that such a thermalization can potentially be avoided either by inducing many-body localizations (MBL) to the system \cite{MBL1,MBL2,MBL3,MBL4,MBL5,MBL6,MBL7} or coupling it to a cold bath \cite{pretherm}. In this case, the robustness of our system under spatial disorders hints the possibility of utilizing the former to combat heating effect if it indeed proves to be detrimental. Moreover, the fact that a physical realization of topological superconductors typically requires proximitizing the system to a normal superconductor provides a natural framework for achieving the latter.

\section{Time induced topology with other periodic drives}
\label{ddf}

As elucidated in Sec.~\ref{intro} of the main text, the nontrivial winding number of the quantity $h_c(t)+\mathrm{i} h_s(t)$ associated with the two periodic drives represents the main mechanism of our construction. As such, it is expected that there exists a class of other time-periodic functions beyond $h_c(t)\propto \cos(\omega t)$ and $h_s(t)\propto \sin(\omega t)$ that is also capable of inducing second-order topology. In particular, given that any time-periodic function $f(t)$ can be Fourier decomposed as $f(t)=\sum_n \left( f^{(s,n)} \sin(n \omega t) + f^{(c,n)} \cos(n\omega t)\right)$, it is generally sufficient to choose the periodically driven hopping amplitudes $J_x(t)$ and $J_y(t)$ to be even and odd in $t$ respectively, so that only $J_x^{(c,n)}$ and $J_y^{(s,n)}$ are nonzero. Note that this choice includes $J_x(t)\propto \cos(\omega t)$ and $J_y(t)\propto \sin(\omega t)$ as a special case.  

\begin{figure}
	\begin{center}
		\includegraphics[scale=0.33]{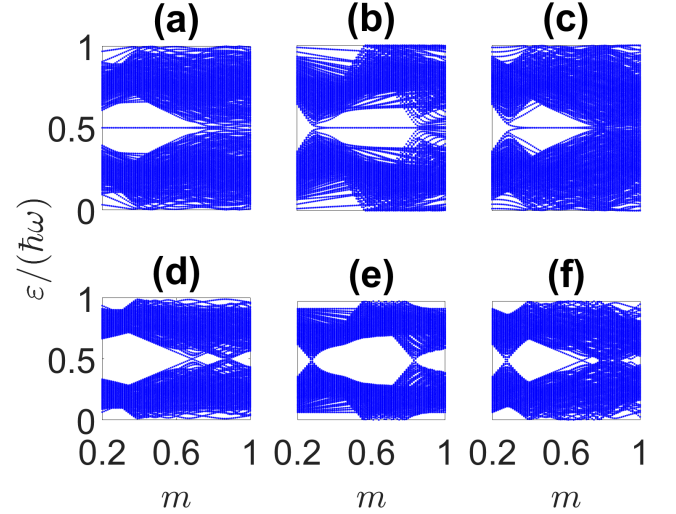}
		\caption{The system's quasienergy spectrum under the modified time-periodicity described in Appendix.~\ref{ddf} where (a,b,c) OBC with $20\times 20$ sites and (d,e,f) PBC are applied in both directions. In panels (a,d), $J_{0,x}=J_{0,y}=2m\hbar$ is varied while $\Delta=\frac{\hbar \omega}{4\pi}$ is fixed. In panels (b,e), $\Delta=\frac{m}{2\pi}\hbar \omega$ is varied while $J_{0,x}=J_{0,y}=0.4\hbar$ is fixed. In panels (c,f), $J_{0,x}=J_{0,y}=\frac{4\pi \Delta}{\omega}=2m\hbar$ is varied. We take $\mu=\frac{0.1}{2\pi}\hbar \omega$ in all panels.}
		\label{pic5}
	\end{center}
\end{figure}  

To provide a concrete example, we may now take $J_x(t)$ and $J_y(t)$ in Eq.~(\ref{sys}) to comprise a series of Dirac delta functions,
\begin{eqnarray}
J_x(t) &=& \sum_\ell J_{0,x} \delta(t-\ell T) \;, \nonumber \\
J_y(t) &=& \sum_\ell J_{0,y} \left(\delta(t-(4\ell+1)T/4)- \delta(t-(4\ell-1)T/4)\right)\;, \nonumber \\ \label{kick}
\end{eqnarray}
which thus include all higher-harmonics in their Fourier decomposition, but $J_x(t)$ ($J_y(t)$) contains only cosine (sine) contributions. In this case, diagonalizing the system's truncated Floquet Hamiltonian no longer represents a feasible way to numerically obtain its quasienergy spectrum as the presence of higher-harmonic terms necessarily requires keeping a large number of Floquet photon sectors to achieve a reasonable accuracy. On the other hand, the one-period time evolution operator of the system under this new driving scheme can be easily obtained as 
\begin{eqnarray}
U_T &=& \exp\left(-\mathrm{i} \frac{H_0 T}{4\hbar}\right)\times \exp\left(\mathrm{i} \frac{H_s}{\hbar} \right) \times \exp\left(-\mathrm{i} \frac{H_0 T}{2\hbar}\right) \nonumber \\
&& \times \exp\left(-\mathrm{i} \frac{H_s}{\hbar} \right) \times \exp\left(-\mathrm{i} \frac{H_0 T}{4\hbar}\right) \times \exp\left(-\mathrm{i} \frac{H_c}{\hbar} \right) \label{Flop} \;, \nonumber \\
H_0 &=& \sum_{i,j} \left(\frac{\mu}{2}c_{i,j}^\dagger c_{i,j}+ \Delta c_{i+1,j}^\dagger c_{i,j}^\dagger +\mathrm{i} \Delta c_{i,j+1}^\dagger c_{i,j}^\dagger +h.c. \right) \;, \nonumber \\
H_s &=& \sum_{i,j} J_{0,y} c_{i,j+1}^\dagger c_{i,j} +h.c. \;, \nonumber \\
H_c &=& \sum_{i,j} J_{0,x} c_{i+1,j}^\dagger c_{i,j}+h.c. \;,
\end{eqnarray}
where $H_s$ and $H_c$ now have units of energy$\times$ time due to the Dirac delta functions. The factorization of $U_T$ into products of six exponentials above can be understood as follows. Within a single period $\left[0,T\right)$, the system's Hamiltonian is constant, except at three times $t_0=0$, $t_1=T/4$, and $t_2=3T/4$ when the Dirac delta terms activate. As a result, the one-period time evolution operator is simply given by the free evolution of $H_0$, interrupted by $H_c$ ($H_s$) for a very short duration at $t_0$ ($t_1$ and $t_2$), which immediately leads to Eq.~(\ref{Flop}). In this case, the system's quasienergies can then be obtained by diagonalizing $U_T$ and taking the phase of its eigenvalues $\exp\left(-\mathrm{i} \varepsilon T/\hbar\right)$.

Under PBC, it can be further verified that the system's Floquet Hamiltonian under the new driving scheme still preserves the four symmetries $\mathcal{P}$, $\tilde{\mathcal{P}}$, $\mathcal{M}_{\rm D}$, and $\mathcal{M}_{\rm AD}$ defined before. Consequently, a similar $Z_2$ invariant can be constructed, i.e., by block anti-diagonalizing the Floquet Hamiltonian in the $\sigma_z$ representation, followed by the calculation of the winding number associated with one block of the anti-diagonal infinite matrices. We will however not pursue this further since its analytical calculation may be more complicated due to additional infinite matrices associated with higher harmonic terms. Instead, we directly evaluate the quasienergy spectrum to demonstrate the presence of corner MPMs in some parameter regime.  

In Fig.~\ref{pic5}, we plot the system's quasienergy spectrum under the modified time-periodic modulations defined by Eq.~(\ref{kick}) as the system parameters are varied. As expected, quasienergy $\frac{\hbar \omega}{2}$ solutions associated with MPMs can be clearly identified for a range of parameter values. In addition, similar to the harmonic driving case with nonzero static hopping amplitudes, chiral MMs around zero quasienergy also exist at some (small) parameter values. On the other hand, we note that more exotic structure is observed at larger parameter values, such as the presence of a topological phase transition between Floquet SOTSC (characterised by the presence of corner MPMs) and anomalous Floquet first-order topological superconductors (characterised by the presence of chiral MMs around $\frac{\hbar \omega}{2}$ quasienergy), as depicted in Fig.~\ref{pic5}(b,e). Such a feature is made possible by the presence of higher-harmonic terms in the new driving scheme. It can thus be envisioned that a variety of topologically nontrivial periodic drives may be utilized to generate novel topological phases displaying other interesting signatures.


\begin{thebibliography}{99}
\bibitem{note0} If $J_{0,x}=J_{0,y}$ and $J_{s,x}=J_{s,y}$, the two symmetries $\mathcal{M}_{\rm D}$ and $\mathcal{M}_{\rm AD}$ correspond to diagonal and anti-diagonal mirror symmetries, i.e., they map $\mathcal{M}_{\rm D} \mathcal{H}'_{\rm BdG}(k_x,k_y)\mathcal{M}_{\rm D}^{(-1)}=-\mathcal{H}'_{\rm BdG}(k_y,k_x)$ and $\mathcal{M}_{\rm AD} \mathcal{H}'_{\rm BdG}(k_x,k_y)\mathcal{M}_{\rm AD}^{(-1)}=-\mathcal{H}'_{\rm BdG}(-k_y,-k_x)$.
\bibitem{note} In real space, particle-hole symmetry maps quasienergy $\varepsilon$ excitation $\Psi_\varepsilon^\dagger$ to $\Psi_{-\varepsilon}$, so that degenerate Hermitian solutions (corresponding to Majorana modes) can only exist at either quasienergy zero (as MZMs) or $\frac{\hbar \omega}{2}$ (as MPMs).
\bibitem{note2} The quasienergy regime $(0,\hbar \omega]$ is used instead of the usual $(-\frac{\hbar \omega}{2},\frac{\hbar \omega}{2}]$ in order to highlight the presence of MPMs at some parameter values.
\end{thebibliography}
\end{document}